\documentclass[12pt,a4paper]{article}

\title{Communication Complexity}
\author{Dömötör Pálvölgyi}
\date{}
\usepackage{amssymb,amsmath,amsthm}
\usepackage[english]{babel}
\usepackage{t1enc}
\usepackage[latin2]{inputenc}
\usepackage{epsfig}

\newtheorem{thm}{Theorem}[section]
\newtheorem{cor}[thm]{Corollary}
\newtheorem{defi}[thm]{Definition}
\newtheorem{lem}[thm]{Lemma}

\newtheorem{claim}[thm]{Claim}

\newtheorem{conj}[thm]{Conjecture}

\usepackage{indentfirst}
\def\dim{\mathrm{dim}}
\def\polylog{\mathrm{polylog}}

\newcommand{\half}{\frac{1}{2}}
\newcommand{\A}{{\bf A }}
\newcommand{\B}{{\bf B }}
\newcommand{\Ax}{{\bf A}}
\newcommand{\Bx}{{\bf B}}

\begin{document}
\thispagestyle{empty} \vspace*{30mm}

\begin{center}
{\fontsize{20.74}{0}\textbf{Communication Complexity 
}}
\vspace{10mm}

{\Large \ \textsc{Thesis} }
\end{center}

{\large \vspace{10mm} }

\begin{center}
{\Large \textsc{ Written by:} \; Dömötör Pálvölgyi }

{\large \vspace{5mm} }

{\Large \textsc{ Supervisor:} \; Zoltán Király }
\end{center}

{\large \vspace{10mm} }

\vskip 5cm
{\large
\begin{figure}[th]
\begin{center}
{\large \epsfig{file=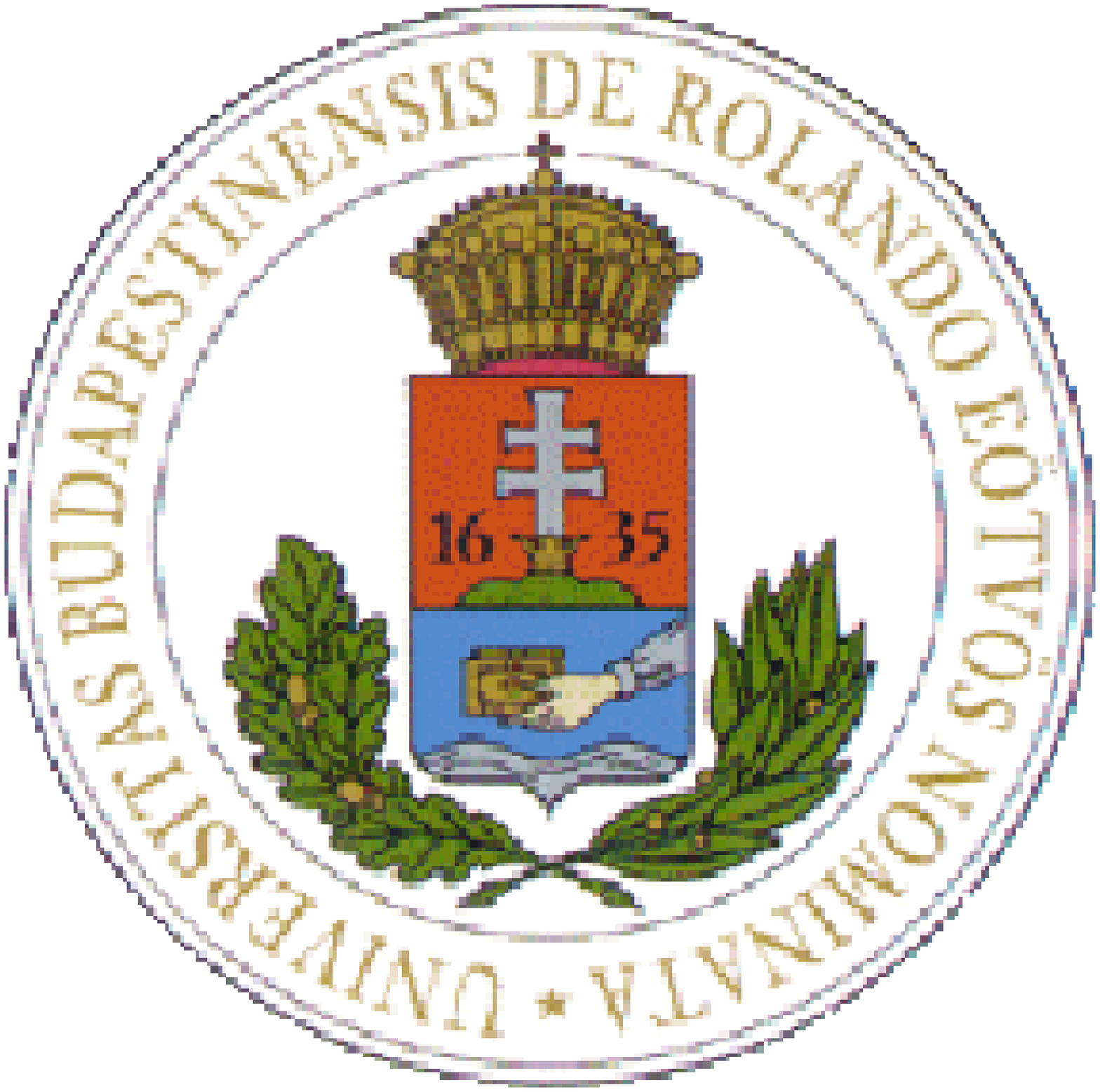,angle=0,width=0.2\textwidth}  }
\end{center}
\par
{\large 
}
\end{figure}
}

\begin{center}
{\large \vspace{10mm} }
{\large Eötvös Loránd University }

{\large Faculty of Sciences }

{\large 2005 }

\end{center}
\newpage
\pagenumbering{arabic}
\tableofcontents

\clearpage
\subsection{The Organization of This Paper}
The first section starts with the basic definitions following mainly the notations of the book written by E. Kushilevitz and N. Nisan \cite{KN96}. This whole thesis started from this book and some parts follow the proofs from there but I always tried to simplify and improve them, I even discovered and corrected a false claim. But most of the results are mine, supervised by Zoltán Király. At the beginning of each part I will indicate if a result is not mine.\\

At the end of the first section I examine tree-balancing. At the very end I introduce a function that I will use only in the later sections.\\

In the second section I summarize the well-known lower bound methods and prove the exact complexity of certain functions.\\

In the first part of the third section I introduce the random complexity and prove the basic lemmas about it. In the second part I prove a better lower bound for the complexity of all random functions. In the third part I introduce and compare several upper bounds for the complexity of the identity function.\\

In the fourth section I examine the well-known \emph{Direct-sum conjecture}. I introduce a different model of computation then prove that it is the same as the original one up to a constant factor. This new model is used to bound the \emph{Amortized Time Complexity} of a function by the number of the leaves of its protocol-tree. After this I examine the Direct-sum problem in case of Partial Information and in the Random case.\\

In the last section I introduce the well-known hierarchy classes, the reducibility and the completeness of series of functions. Then I define the class \emph{PSPACE} and \emph{Oracles} in the communication complexity model and prove some basic claims about them.\\

There are plenty questions left open in this paper, I hope I can
manage to make some progress in some of them during the
forthcoming years.\\

I would like to thank Zoltán Király, my supervisor, for his guidance, help and inspiration. This thesis would be nowhere without him.

\clearpage
\section{Introduction and Tree-balancing}

\subsection{Introduction}
Communication Complexity is a computational model introduced by Yao in 1979 \cite{Y79}. Since then, many papers have been written about it, more likely because of its consequences in applications, including VLSI theory, and because it is more tractable than computational complexity, it is easier to prove lower bounds. It also has a very simple definition.\\

Two players, commonly denoted by \A and \B and called in many names from Alice and Bob through Alfred and Paul to Rosencrantz and Guildenstern, both hold an input from a finite set, $x \in X$ and $y \in Y$, and wish to compute $f(x,y)$ where $f$ is a given function known to both of them. To do this, of course, they have to communicate. They are allowed to send bits to each other. For example, we give both of them a number from $1$ to $10$ and ask them whether their numbers are the same or not. They can solve this, if \A transforms her number to base two numbering system, then she sends each bit to \Bx , who can compare it with his number and send the result back to her. This algorithm requires 4 + 1 bits.\\

The last bit of their communication have to be the value of the function if the range of the function is $\{0,1\}$, this is why in the case of determining the exact length of the necessary communication, we write it with a '+1' to avoid confusion. Another reason to use this notation is that if only one of them has to know the answer, then they have to communicate exactly one bit less. When we are interested only in the order of magnitude, sometimes we even omit constant multipliers.\\

A more challenging task, as usual, is to prove a lower bound for the length of the worst case communication. In this paper we present several techniques for this.

\subsection{Basic Definitions}
In this section we follow the book (\cite{KN96} pp. 3-10 and 16-19).

The formal definition of Communication Complexity is the following:

\begin{defi} For a function $f: X \times Y \rightarrow Z$, the (deterministic) communication complexity of $f$ is the minimum length of \bfseries P\mdseries , over all protocols \bfseries P \mdseries that compute $f$. We denote it by $D(f)$.
\end{defi}

Of course, we have to define $Protocol$:
\begin{defi} A Protocol is a binary tree. There is a function $X \rightarrow \{0,1\}$ or $Y \rightarrow \{0,1\}$ associated to each of the internal nodes and there is an element from $Z$ belonging to each leaf. If we start from the root of the tree and turn at each node at the direction determined by the function, we have to end in a leaf with the value $f(x,y)$.
\end{defi}

It is easy to see that this is equivalent to the communication of \A and \B as described in the previous section. In the proofs we are often using this fact because it is easier to imagine it like that.\\

In the following claim and throughout this paper, let $\log n$ denote $\log_2 n$, although sometimes we mean $\left\lceil \log_2 n \right\rceil$ or $\left\lfloor \log_2 n \right\rfloor$, the reader may always easily figure out which.

\begin{claim} \label{logZ} $D(f) \leq \log \left| X \right| + \log \left| Z \right|$
\end{claim}
\begin{proof}
\A sends a 0-1 sequence that encodes her input (in a predetermined way), then \B computes the value of $f$ using his unlimited computational power and sends back the result to her in a 0-1 sequence.
\end{proof}

From now on, $Z$ will be $\{0,1\}$ because these are the most studied functions and the ones we are dealing with in this paper. We also assume, unless we state otherwise, that $X = Y = \{0,1\}^n$.\\

We can imagine $f$ as a matrix; the rows are representing the inputs of \A while the columns are representing the inputs of \Bx . The entries are the values associated to the corresponding inputs.\\

Now we define some of our favorite functions:

\begin{defi} $ $
\begin{itemize}
    \item $EQ(x,y) = 1$ iff $x = y$.
    \item $NE(x,y) = 0$ iff $x = y$.
    \item $GT(x,y) = 1$ iff $x \geq y$.
    \item $IP(x,y) =$ $\left\langle x,y\right\rangle\mod 2= \sum_i x_iy_i \mod 2$.
    \item $DISJ(x,y) = 1$ iff $\sum_i x_iy_i = 0$.
    \item $\bar{F}(x,y) = 0$ iff $F(x,y)=1$.
\end{itemize}
\end{defi}

Because of Claim \ref{logZ}, $D(f) \leq n + 1$ holds for all $f$s whose range is $\{0,1\}$, thus for all of the above functions.\\

It is easy to see that in each node of the protocol-tree one of the players, whose turn it is to speak at that node, splits one's set of inputs into two parts. This is equivalent to dividing the set of rows (or columns) into two. So after each step, we have another communication problem to solve, that has a smaller matrix formed by the remaining rows and columns of the players.

\begin{defi} A set $R$ of entries is a rectangle in a matrix iff $R = I \times J$ where $I \subseteq X$ is a set of rows and $J \subseteq Y$ is a set of columns. A rectangle is monochromatic if all of its entries are the same. We call these respectively 0-rectangle and 1-rectangle.
\end{defi}

\begin{claim} The input-pairs leading to the same leaf in a protocol form a monochromatic rectangle. Moreover, these rectangles partition the matrix.
\end{claim}
\begin{proof}
We prove by induction that in each step of the protocol the input-pairs leading to that node form a rectangle. This is true since in each node of the protocol-tree one of the players is splitting her remaining inputs into two disjoint sets, both giving a rectangle because of the induction. The rectangle of the leaf is monochromatic because the protocol computes $f$.\\
Each entry belongs to exactly one leaf, thus we get a partition indeed.
\end{proof}

We call the leaves that lead to a 0-rectangle 0-leaves and those that lead to a 1-rectangle 1-leaves.\\

Note that not all rectangle-partitions can be associated to protocols.
To distinguish, we need the following Definition:

\begin{defi} $ $
\begin{itemize}
    \item The protocol partition number of $f$, $C^P(f)$, is the smallest number of leaves in a protocol-tree that computes $f$.
    \item The protocol 0-partition number of $f$, $C_0^P(f)$, is the smallest number of 0-leaves in a protocol-tree that computes $f$.  Similarly, $C_1^P(f)$ is the smallest number of 1-leaves.
    \item The partition number of $f$, $C^D(f)$, is the smallest number of monochromatic rectangles that can partition the matrix of $f$.
    \item The 0-partition number of $f$, $C_0^D(f)$, is the smallest number of 0-rectangles in a monochromatic rectangle partition of the matrix of $f$. Similarly, $C_1^D(f)$ is the smallest number of 1-rectangles.
    \item The cover number of $f$, $C(f)$, is the smallest number of monochromatic rectangles that cover all the entries of the matrix of $f$.
    \item The 1-cover number of $f$, $C_1(f)$, is the smallest number of 1-monochromatic rectangles that cover the 1 entries of the matrix of $f$. Similarly, $C_0(f)$, is the smallest number of 0-monochromatic rectangles that cover the 0 entries of the matrix of $f$.
\end{itemize}
\end{defi}

The following inequalities obviously follow:

\begin{claim} $C(f) \leq C^D(f) \leq C^P(f)$,\\
$C_0(f) \leq C_0^D(f) \leq C_0^P(f)$,\\
$C_1(f) \leq C_1^D(f) \leq C_1^P(f)$,\\
$C_0^P(f) + C_1^P(f) \leq C^P(f)$,\\
$C_0^D(f) + C_1^D(f) = C^D(f)$,\\
$C_0(f) + C_1(f) = C(f)$.
\end{claim}

We can give a lower bound for $D(f)$ by the logarithm of any of these values. This is the corollary of the previous and the following claim:

\begin{claim} \label{DC1} $D(f) \geq 1 + \log C_1^P(f)$.
\end{claim}
\begin{proof}
We split each leaf that is not at the bottom (the deepest part) of the tree into a 1-leaf and a 0-leaf. Now the depth did not increase and we have at least $2C_1^P(f)$ leaves. A binary tree with depth $D(f)$ can have at most $2^{D(f)}$ leaves, so we have $2^{D(f)} \geq 2C_1^P(f)$, just what we wanted.\\
\end{proof}

\begin{cor} \label{DC0} $D(f) \geq 1 + \log C_0^P(f)$.
\end{cor}

\begin{cor} $D(f) \geq \log C_P(f)$.
\end{cor}

The most important of these values is $C^P(f)$ because $D(f) = \Theta (\log C^P(f))$. This can be proved with a tree balancing Lemma.

\subsection{Tree-balancing}
Here the first lemma is from the book (\cite{KN96} pp. 19-20), the other one is our result.\\

\begin{lem} \label{CpD} $\log C^P(f) \leq D(f) \leq 3 \log C^P(f)$.
\end{lem}
\begin{proof}
The first inequality follows from the previous claim.\\
To prove the second one, we have to construct a shallow protocol from a protocol-tree with $C^P(f)$ leaves. We prove by induction on the number of leaves that $D(f) \leq 3 \log C^P(f)$.

\begin{claim} In a tree there always exists a vertex, such that leaving out this vertex, all components of the remaining graph have at most half of the original leaves.
\end{claim}
\begin{proof}
If an edge splits the leaves into two even parts, we are done. Otherwise, direct each edge toward the bigger part. There must be a node with outdegree 0, this will do.
\end{proof}

We choose such a node in our protocol-tree. \A and \B both send a bit indicating whether their inputs allow the path to the node, that means whether their input intersects the rectangle of that node. (We say that an input intersects a rectangle, if the row (column) of the input is the row (column) of the rectangle.) If one of their inputs does not intersect the rectangle, we leave out this node and everything under it, we have halved the number of nodes with 2 bits. If both of their inputs intersect the rectangle, then we can restrict the rest of the game only to this rectangle. The person who has to speak at this node sends one more bit, now we have halved the number of leaves (because of the splitting property of the chosen node) with 3 bits. Thus we can halve the number of leaves in both cases with at most 3 bits of communication, this completes our proof.
\end{proof}

The exact relation between $C^D(f)$ and $C^P(f)$ is yet unsolved although Kushilevitz et al. \cite{KLO96} showed that a small gap ($2 C^D(f) = C^P(f)$) is possible.\\

We have another useful lemma about balancing protocol-trees that does not hold in most computational models but it does in communication complexity. We can prove that $C_1^P(f)$ and $C_0^P(f)$ can differ only in a constant factor, so communication complexity is \textit{result-balanced}, we have almost the same number of both type of outcomes. We prove this here using only certain local transformations of the protocol-tree, this gives the following result:

\begin{lem} \label{4} $C_0^P(f) \leq 4C_1^P(f) - 2$ unless $f$ is constant.
\end{lem}
\begin{proof}
In fact we prove a somewhat stronger statement, that this holds for any protocol-tree without unnecessary nodes. Let $T$ be an arbitrary protocol-tree. We denote by $L(T)$ the number of leaves, $L_1(T)$ ($L_0(T)$) denotes the number of 1-leaves (0-leaves). We denote by $T_z$ the tree that we get by chopping the tree $T$ at its node $z$, so the nodes of the new tree are $z$ and its descendants. We denote by $z_l$ and $z_r$ the children of $z$. We prove by induction on $L(T)$. In fact we are going to use that $L_0(T_z) \leq 4L_1(T_z) - 2$ holds for all $z$ non-leaf descendants of the node $v$ and from this we obtain $L_0(T_v) \leq 4L_1(T_v) - 2$ or we find an unnecessary node. For any node $z$ we have one of the following cases:\\

(1) Both of $z_l$ and $z_r$ are leaves; one of them have to be a 0-leaf, the other a 1-leaf or $z$ would be unnecessary. This implies $L_0(T_z) \leq 4L_1(T_z) - 3$.\\

(2) None of $z_l$ and $z_r$ is a leaf; from the induction we have $L_0(T_{z_l}) \leq 4L_1(T_{z_l}) - 2$ and $L_0(T_{z_r}) \leq 4L_1(T_{z_r}) - 2$. This implies $L_0(T_z) \leq 4L_1(T_z) - 4$.\\

(3) One of $z_l$ and $z_r$ is a 1-leaf; the induction on the other child implies $L_0(T_z) \leq 4(L_1(T_z) - 1) - 2 = 4L_1(T_z) - 6$.\\

(4) One of $z_l$ and $z_r$ is a 0-leaf.\\

Now we can follow a straightforward argument.\\

If for $v$ we have one of the first three cases, we are done. So the only interesting case is when one of the children of $v$ is a 0-leaf while the other is not a leaf. Wlog, we can suppose that at $v$ it is \Ax 's turn to speak and we denote the non-leaf child of $v$ by $w$.\\

\begin{figure}[!h]

\bigskip\bigskip

\begin{center}
{ \epsfig{file=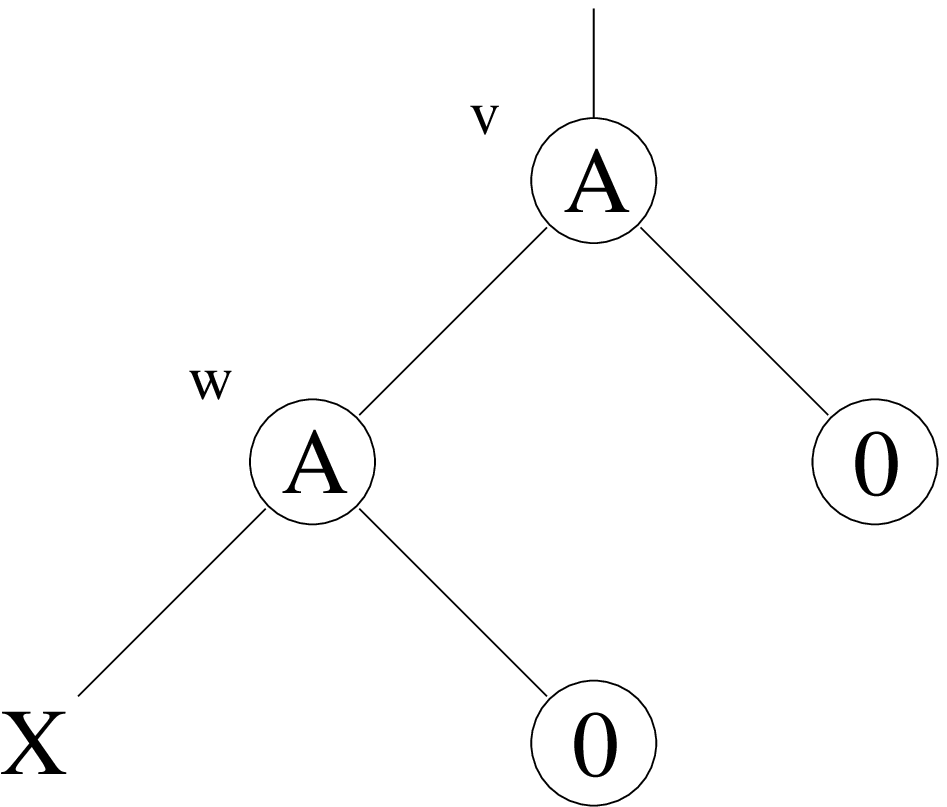,angle=0,width=0.3\textwidth} } ~~~~~~~~~~~~~~~~~~~~~~~~~~~
{ \epsfig{file=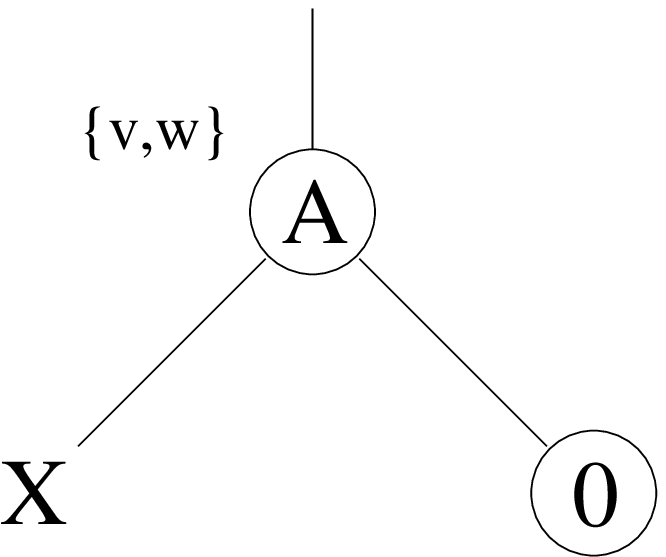,angle=0,width=0.22\textwidth} }

\bigskip

Before ~~~~~~~~~~~~~~~~~~~~~~~~~~~~~~~~~~~~~~~~~~~~~~~~~~~~~ After

\end{center}
\par

\bigskip

\bf\caption{\normalfont Trivial reduction} \label{triv}

\bigskip

{\large
}
\end{figure}

Again, for $w$ we must have the fourth case or we are done. If at $w$ it were also \Ax 's turn to speak, then she could merge $v$ and $w$, so one of the nodes is unnecessary. (See {\bf Figure \ref{triv}.}) So at $w$ it is \Bx 's turn to speak. We shall denote the non-leaf child of $w$ by $u$.\\

If for $u$ we have the second or the third case, we are done. Both in the first and the fourth case one of $u$-s children is a 0-leaf. If it is \Bx 's turn to speak, he could merge $w$ and $u$ (like in {\bf Figure \ref{triv}} $v$ and $w$ were merged by \Ax ), hence we had an unnecessary node. If it is \Ax 's turn to speak, then she could merge $v$ and $u$, and \B can speak what he had to at $w$ after this node, hence again one of the nodes would be unnecessary. (See {\bf Figure \ref{trivskip}.})\\

\begin{figure}[th]

\bigskip\bigskip

\begin{center}
{ \epsfig{file=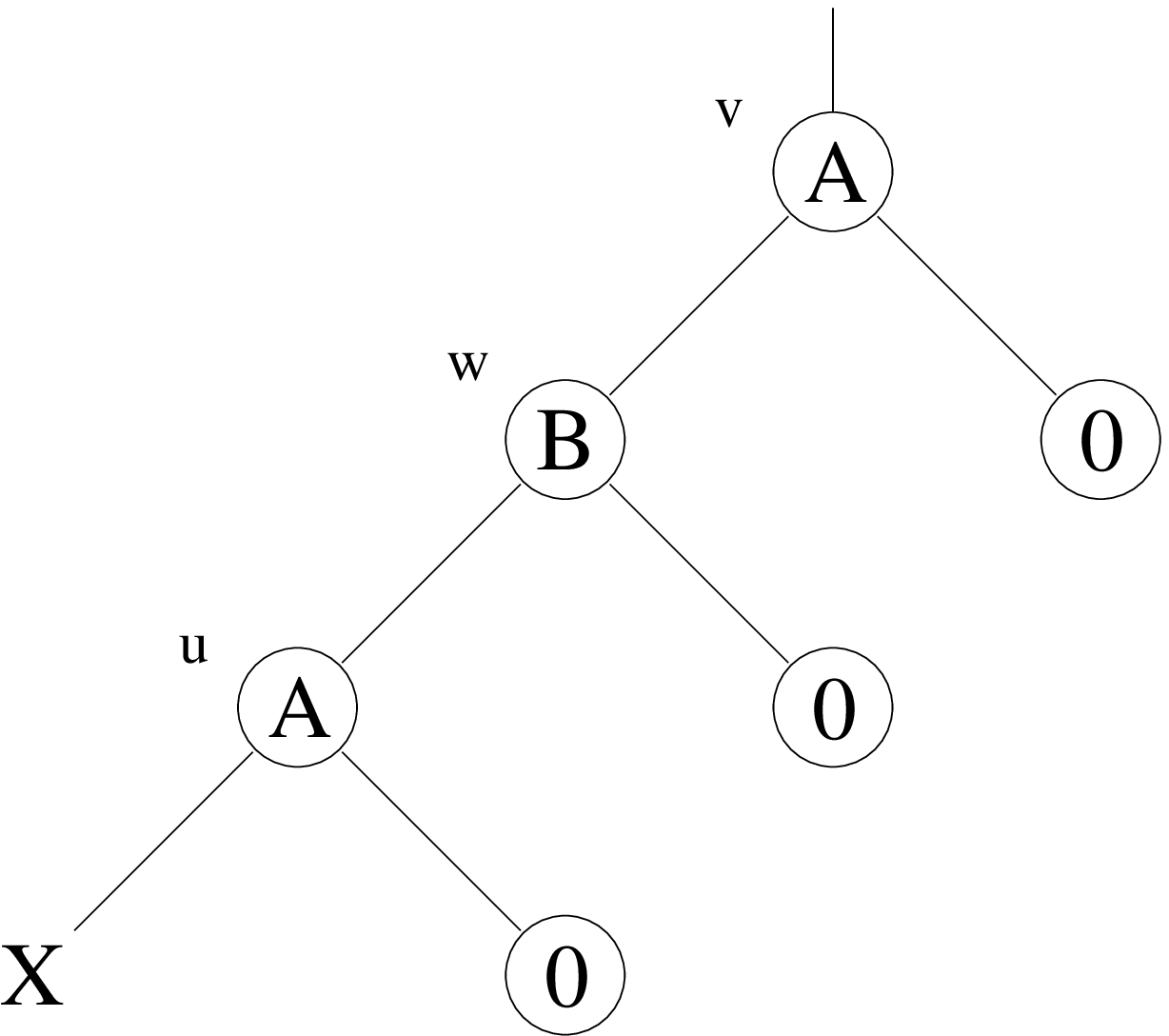,angle=0,width=0.4\textwidth} } ~~~~~~~~~~~~~~~~~
{ \epsfig{file=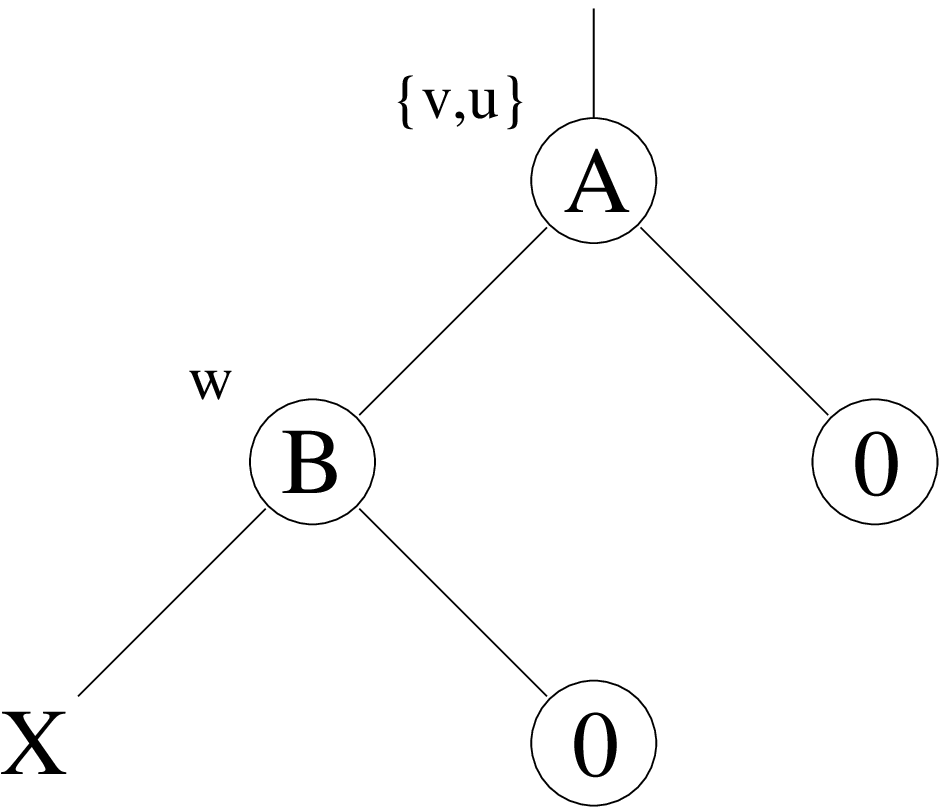,angle=0,width=0.3\textwidth} }

\bigskip

Before ~~~~~~~~~~~~~~~~~~~~~~~~~~~~~~~~~~~~~~~~~~~~~~~~~~~~~~~ After

\end{center}
\par

\bigskip

\bf\caption{\normalfont Reduction with one skip} \label{trivskip}

\bigskip

{\large
}
\end{figure}

This completes the proof, so we gained $L_0(T_v) \leq 4L_1(T_v) - 2$. This holds for all nodes, for the root of tree as well, so we have proved $C_0^P(f) \leq 4C_1^P(f) - 2$.
\end{proof}

Of course, we can switch the roles of 0- and 1-leaves.

\begin{cor} $C_1^P(f) \leq 4C_0^P(f) - 2$ unless $f$ is constant.
\end{cor}

We also obtained the following sharpening of Claim \ref{DC1}:

\begin{cor} $D(f) = \Theta(C_1^P(f))$.
\end{cor}

Note that none of the above lemmas is sharp, we do not know the exact values. For the latter, we can prove a sharper upper bound, that gives $3/2$ instead of $4$. This is done in the following section.

\subsection{Advanced Tree-balancing}
On a protocol-tree $P$ that computes an arbitrary function we are going to make certain transformations that increase neither the number of 0-leaves, nor the number of 1-leaves, and do not even increase the depth of the tree. If none of these transformations can be made on $P$, then we can prove that the number of 1-leaves is very close to the number of 0-leaves. Of course, this implies the same for $C_1^P(f)$ and $C_0^P(f)$.\\

We call a node \Ax -node if at the node it is \Ax 's turn to speak. We call a leaf \Ax -leaf if at the father of the leaf it is \Ax 's turn to speak. We similarly define \Bx -node and \Bx -leaf.\\
The \emph{magnitude} of a leaf is the number of leaves under its father. Hence it is always at least two.\\

\begin{figure}[th]

\bigskip\bigskip

\begin{center}
{ \epsfig{file=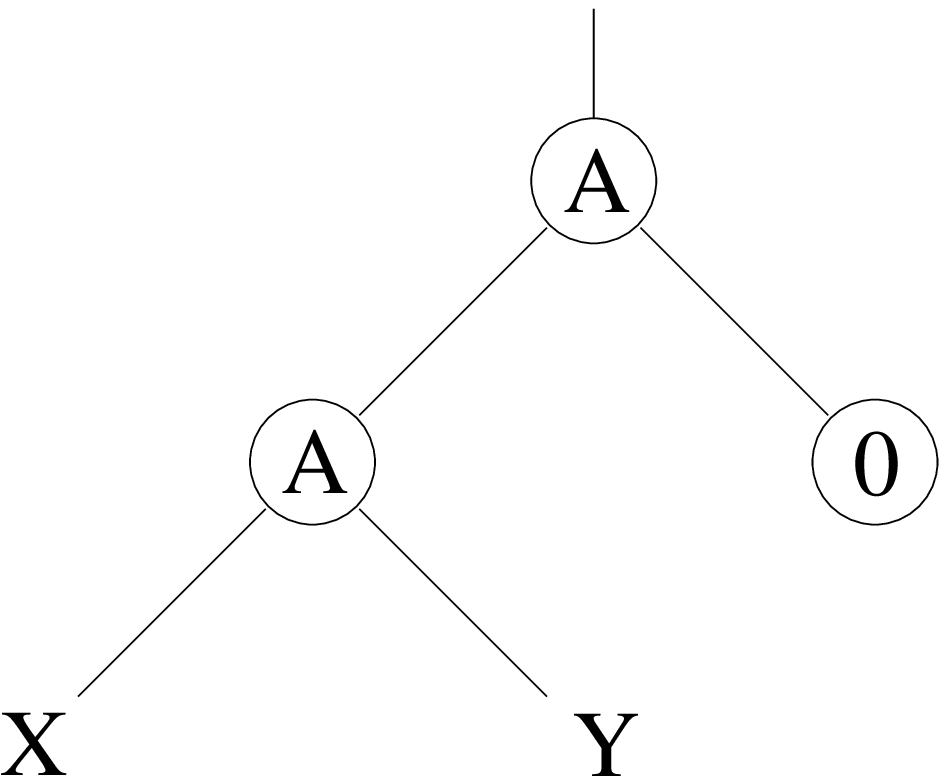,angle=0,width=0.31\textwidth} } ~~~~~~~~~~~~~~~~~
{ \epsfig{file=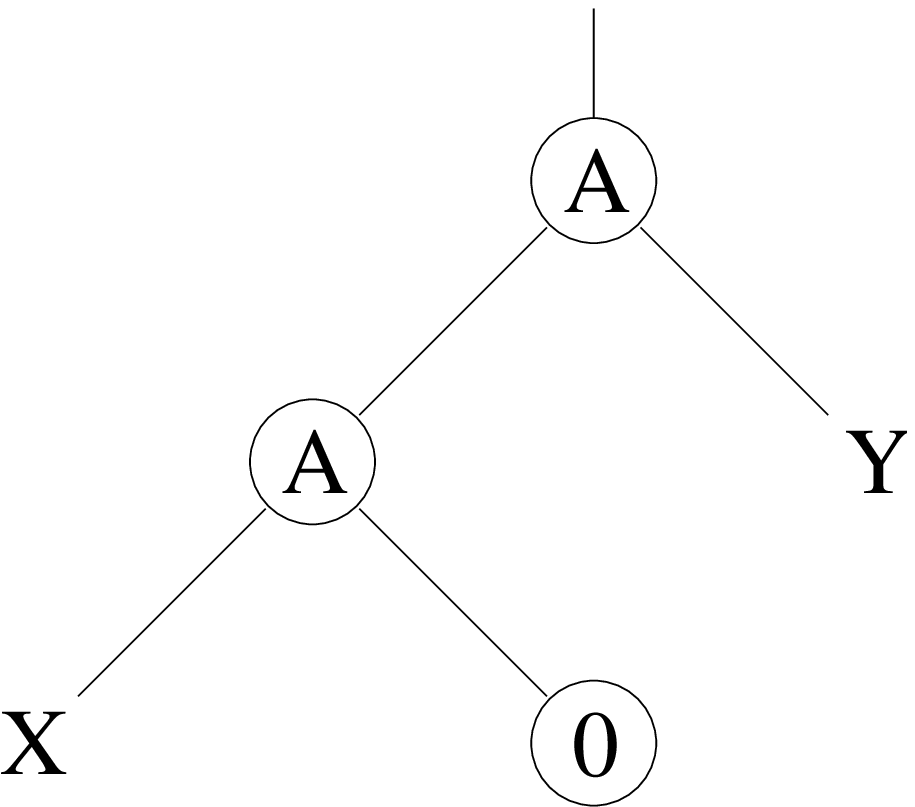,angle=0,width=0.3\textwidth} }

\bigskip

Before ~~~~~~~~~~~~~~~~~~~~~~~~~~~~~~~~~~~~~~~~~~~~~~~~~~~~~~~ After

\end{center}
\par

\bigskip

\bf\caption{\normalfont Push down through own node} \label{pushdown}

\bigskip

{\large
}
\end{figure}

The idea is to \emph{push down} the 0-leaves in the tree as deep as possible. Eg., if an \Ax -node has a 0-leaf hanging from it and its other son is an \Ax -node as well, we can switch the two nodes. (See {\bf Figure \ref{pushdown}.}) So we can assume that if a 0-leaf is hanging down from a node, then at the other son of that node it is the other person's turn to speak.\\

\begin{figure}[th]

\bigskip\bigskip

\begin{center}
{ \epsfig{file=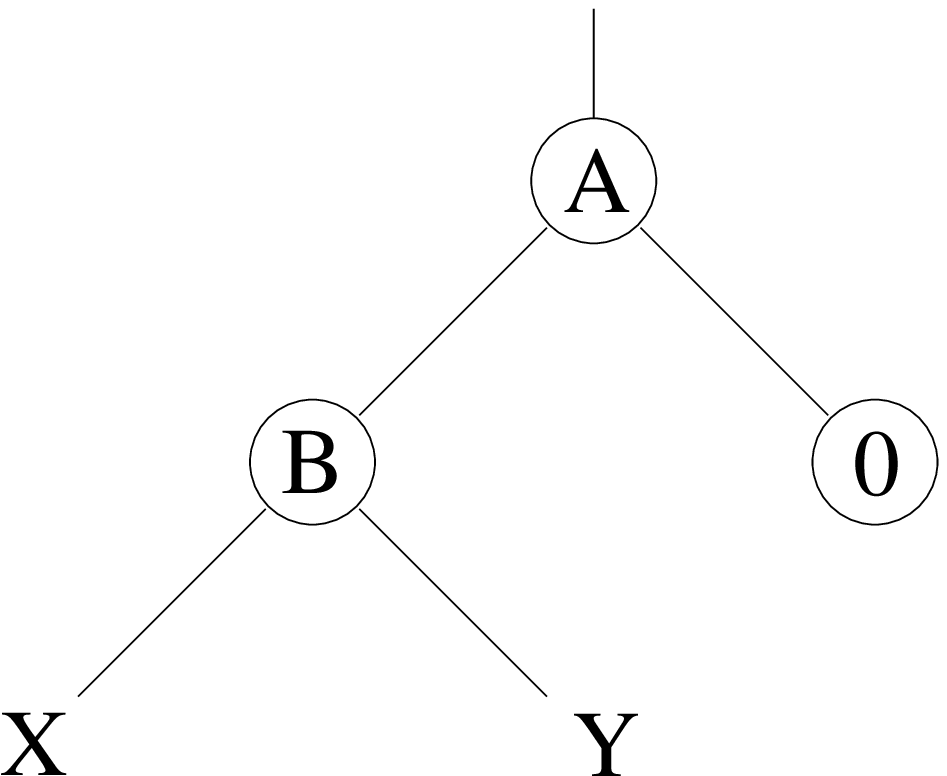,angle=0,width=0.31\textwidth} } ~~~~~~~~~~~~~~~~~
{ \epsfig{file=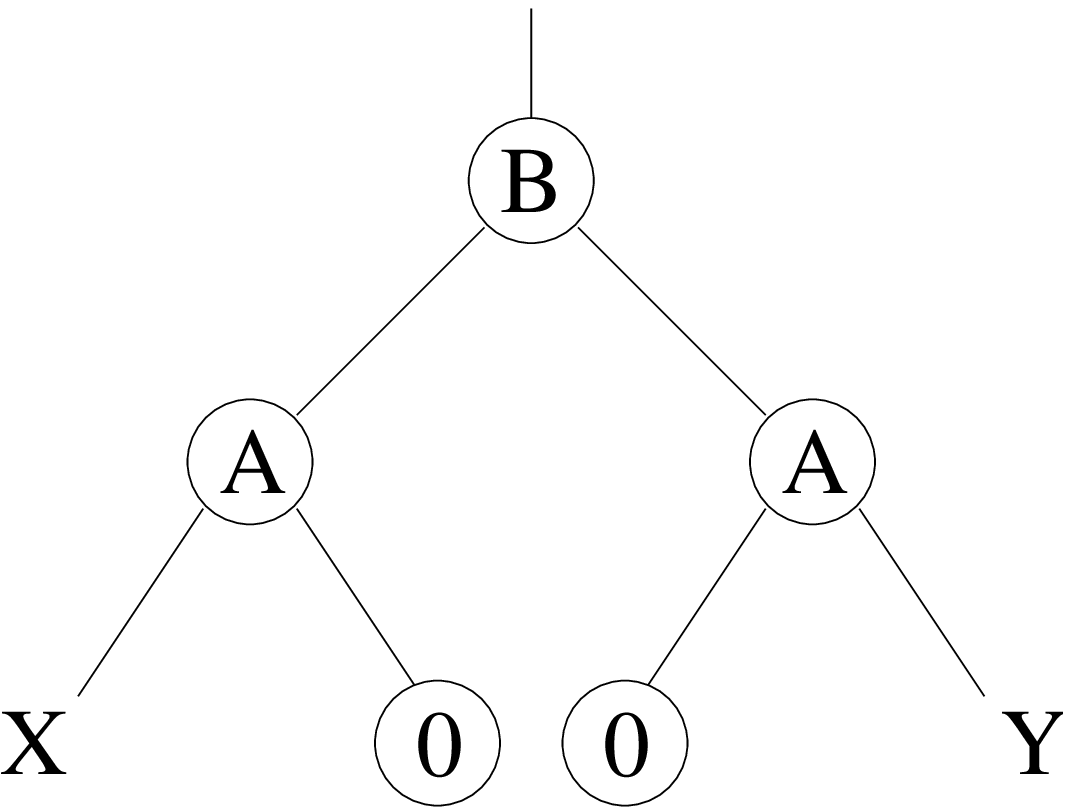,angle=0,width=0.35\textwidth} }

\bigskip

Before ~~~~~~~~~~~~~~~~~~~~~~~~~~~~~~~~~~~~~~~~~~~~~~~~ After

\end{center}
\par

\bigskip

\bf\caption{\normalfont Push down through other's node} \label{pushdownB}

\bigskip

{\large
}
\end{figure}

Can we push down a 0-leaf even deeper? Let us see what happens if we switch the father of an \Ax -0-leaf with its son! Now unfortunately a new 0-leaf was created (see {\bf Figure \ref{pushdownB}}) but the depth of the tree did not increase. We keep trying to push down these 0-leaves, sometimes duplicating them again when pushing through a node of \Bx . Some get eliminated (when meeting another 0-leaf), the others reach the bottom of the tree. The last node that they pass must be a \Bx -node with a 1-leaf, otherwise we could push our leaf even deeper or eliminate our 0-leaf with an other 0-leaf. If at most one 0-leaf reached the bottom, this transformation did not increase the number of 0-leaves, thus we can execute it. If at least two 0-leaves would reach the bottom, we do not perform the transformation but we associate the \Bx -1-leaves that we have reached with our original \Ax -0-leaf. Note that we do not associate a \Bx -1-leaf to two different \Ax -0-leaves; the path leading from the higher one would go through the parent of the other but it is impossible because there it should have been eliminated.\\

We perform the above operations as many times as we can. We would like to prove that we do not get into an infinite loop; we need something that strictly decreases after each transformation. If a leaf was eliminated, we have no problem, the number of leaves decreased. If not, let us consider the sum of the \emph{magnitudes} of 0-leaves.\\ 

\begin{figure}[th]

\bigskip\bigskip

\begin{center}
{ \epsfig{file=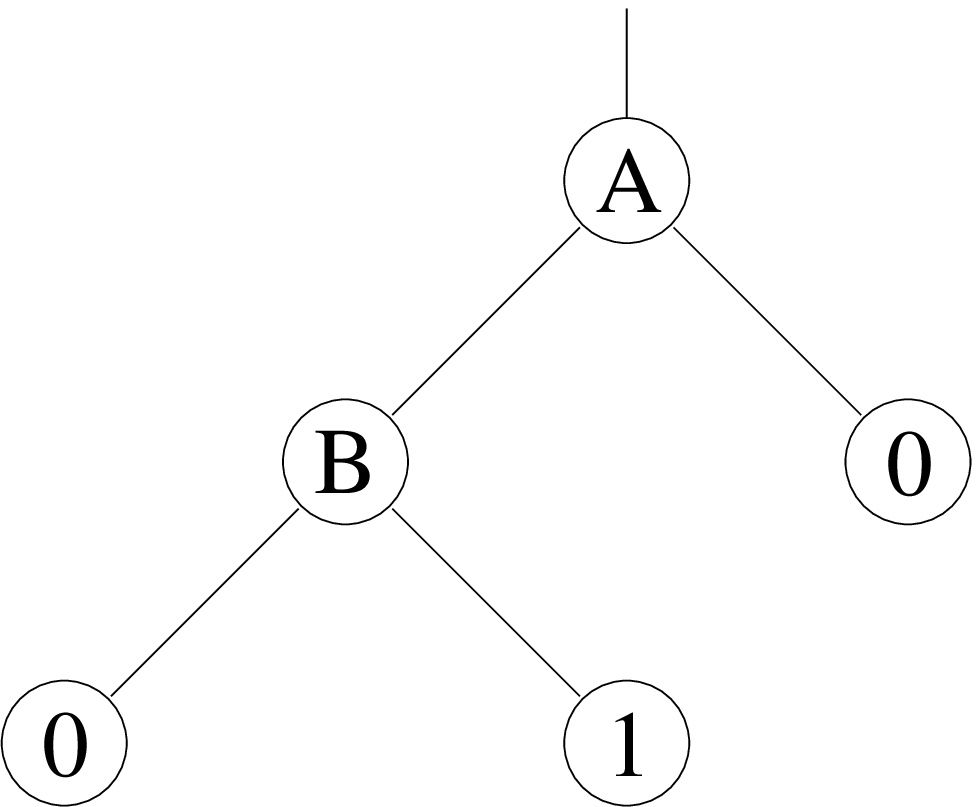,angle=0,width=0.31\textwidth} } ~~~~~~~~~~~~~~~~~
{ \epsfig{file=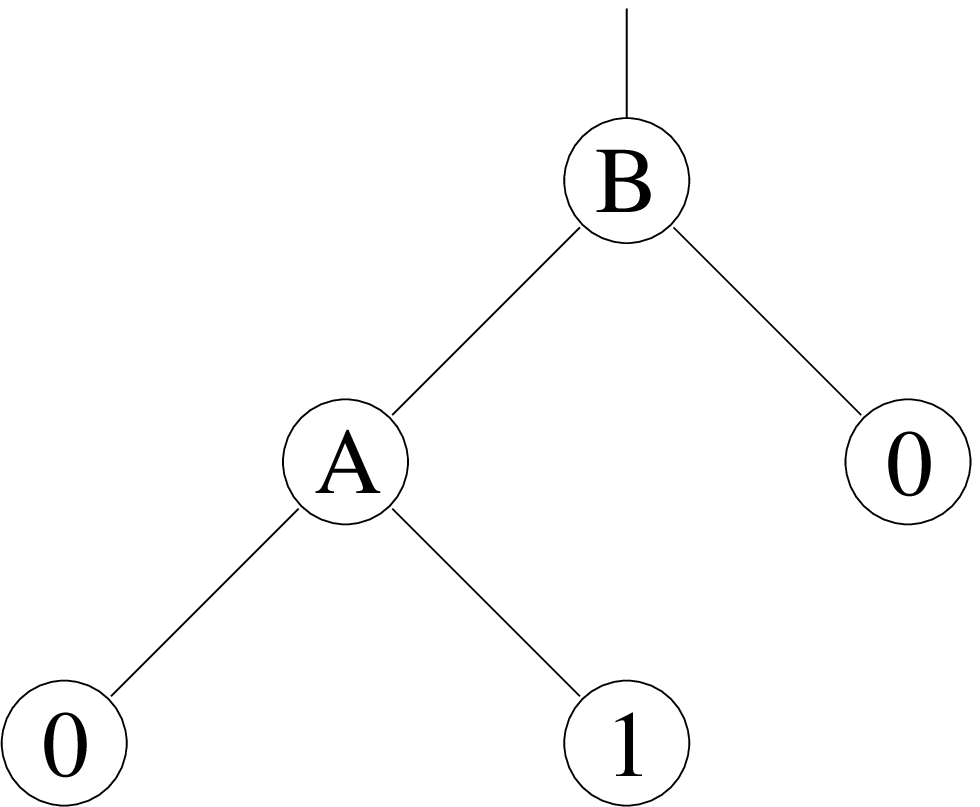,angle=0,width=0.31\textwidth} }

\end{center}
\par

\bigskip

\bf\caption{\normalfont The Twists} \label{twist}

\bigskip

{\large
}
\end{figure}

Wlog, let us assume that an \Ax -0-leaf is being pushed down. During the push down through an \Ax -node, this sum will strictly decrease; its only element that changes is the magnitude of our \Ax -0-leaf, it will decrease by the number of leaves in $Y$. (See {\bf Figure \ref{pushdown}.}) During the push down over a \Bx -node, this sum cannot grow; only one of the two new 0-leaves survives and it has fewer 0-leaves under it than our original \Ax -0-leaf had. However, it is possible that this sum remains the same, if one of the children of the \Bx -node is a 0-leaf. But in this case, we can push down either this \Bx -0-leaf or our new \Ax -0-leaf through the other son of the \Bx -node (depending whether it is a \Bx - or an \Ax -node) unless it is a 1-leaf. This is the only configuration that is wrong. (See {\bf Figure \ref{twist}.}) We are going to call this configuration and the one we can obtain from it by switching the node of \A and \Bx , a \emph{twist} and denote it by $T$. (These two can be obtained from each other, hence we do not need to distinguish them.)\\

If we encounter a twist, we do not perform the transformation. Therefore we cannot get into an infinite loop.\\

If we can perform none of the above operations, it means that every \Ax -0-leaf whose brother is not a 1-leaf, and who is not in a twist, has at least two \Bx -1-leaves associated to it. We will call the paths leading from the father of the \Ax -0-leaf to the \Bx -1-leaves \Ax -paths. We can similarly define the \Bx -paths.\\

It is high time to introduce some notations:\\
\\
We denote the number of twists by $T$.\\
We denote the number of \Ax -0-leaves by $a_0$.\\
We similarly denote the number of \Ax -1-leaves by $a_1$, the number of \Bx -0-leaves by $b_0$ and the number of \Bx -1-leaves by $b_1$.\\
We denote the number of \Ax -0-leaves who are not in a twist and whose brother is also a leaf by $a_0^{end}$. We denote the number of \Ax -0-leaves who are not in a twist and whose brother is not a leaf by $a_0'$. We similarly define this for the other type of leaves as well. (Note that in each twist we have an \Ax -0-leaf, a \Bx -0-leaf and a 1-leaf, and none of them belongs to the $^{end}$ or to the $'$ part. The 1-leaf can belong to any of the players, we can arbitrarily switch the nodes of the twist as we want to, so it makes no difference where we count the 1-leaf.)\\
If we do not want to specify whether it is an \Ax - or a \Bx -leaf, we denote the corresponding number by $L$ indexed by the appropriate things. Eg., $a_1 + b_1 = L_1$.\\

Hence we have $T + a_0^{end} + a_0' = a_0$, $T + b_0^{end} + b_0' = b_0$ and $T + a_1^{end} + a_1' + b_1^{end} + b_1' = a_1 + b_1$.\\
Because of the associations, we have $b_1 \geq 2a_0'$ and $a_1 \geq 2b_0'$. Moreover, it is easy to see that an \Ax -path or a \Bx -path can not lead to a twist-leaf. (If it did, the 0-leaf would be eliminated by the 0-leaves of the twist.) Hence $a_1 + b_1 - T \geq 2a_0' + 2b_0'$.\\
We also know that $a_1^{end} = a_0^{end}$ and $b_1^{end} = b_0^{end}$ unless $f$ is constant, but let us exclude this trivial case.\\
Putting these all together we get $a_0 + b_0 = T + a_0^{end} + a_0' + T + b_0^{end} + b_0' \leq 2T + \half(a_1 + b_1 - T) + a_1^{end} + b_1^{end} = \half L_1 + \frac{3}{2} T + L_1^{end} \leq \half L_1 + \frac{3}{2} (T + L_1^{end}) \leq 2 L_1$. This immediately leads to the following better improvement of Lemma \ref{4} for all nonconstant $f$s:

\begin{lem} $C^P_0(f) \leq 2 C^P_1(f)$ unless $f$ is constant.
\end{lem}

Now all we have to do is reduce that constant $\frac{3}{2}$ in front of $T$ to $1$ to obtain the desired upper bound. Note that this does not hold if the whole protocol-tree is a single twist. However, miraculously we can use the twists to prove the bound for every other tree. So let us suppose that our protocol-tree has more than four leaves, so it is not constant or a single twist.\\

We are going to associate a non-twist 1-leaf that is not associated to any 0-leaf to each twist. Let us fix a twist, $T^*$. We can suppose wlog, that the father of $T^*$ is an \Ax -node. In this case there can be no \Ax -path through this node. (If there were, the \Ax -0-leaf could be eliminated by turning into the twist at its father.) Furthermore, we can start a \emph{twist-path} going down from the father of $T^*$; it is equivalent to hypothetically moving the \Ax -0-leaf out of the twist (switching its father with the father of the twist) and then moving it down, exactly how we did with an ordinary \Ax -0-leaf with the push down operations when we created the \Ax -paths. The only exception is that we do not move it deeper; we only check whether it can be eliminated or not. If it can be, we reduced the number of 0-leaves, we are happy. If not, then we can associate it to a non-twist, so-far-not-associated \Bx -1-leaf.\\

This gives us the following inequality: $T + 2a_0' + 2b_0' \leq a_1 + b_1 - T$ if $L > 3$.\\
This leads to the improved inequality $a_0 + b_0 \leq \half L_1 + T + L_1^{end} \leq \frac{3}{2} L_1$ if $L > 3$.\\
So we can state our final version of the result-balancing lemma:

\begin{lem} $C^P_0(f) \leq \frac{3}{2} C^P_1(f)$ unless $C^P(f) \leq 3$.
\end{lem}

\begin{cor} $C^P_1(f) \leq \frac{3}{2} C^P_0(f)$ unless $C^P(f) \leq 3$.
\end{cor}

However, it still remains an open question whether the $\frac{3}{2}$ can be improved or not. Even $C^P_0(f)/C^P_1(f) \stackrel{C^P(f) \rightarrow \infty}{\longrightarrow} 1$ is possible, but we do not think that is likely.

\subsection{A Technical Function}
Here we define a real function that comes up often in this paper and we give an estimation for it. We denote the inverse of a function by $^{(-1)}$.

\begin{defi} $\Lambda(s) = 2^{s + \log s} = s2^s$, $\lambda(t) = \Lambda^{(-1)}(t)$.
\end{defi}

The estimation easily follows from the definition:

\begin{claim} \label{L-est} $\log t - \log \log t \leq \lambda(t) \leq \log t$.
\end{claim}

In fact with the help of this $\lambda$ function, we can get a formula for any similar function:

\begin{claim} $(s + a)2^s = 2^{-a}\Lambda(s + a)$.
\end{claim}
\begin{proof}
$\Lambda(s + a) = (s+a)2^{s + a} = 2^a(s + a)2^s$.
\end{proof}

\begin{cor} \label{L-shift} $((s + a)2^s)^{(-1)}(t) = \lambda(t2^a) - a$.
\end{cor}

\clearpage
\section{Lower Bounds}
There are three general methods to give a good lower bound for $D(f)$. This whole section follows the book (\cite{KN96} pp. 10-14) except Claim \ref{IP1} and the proving of the exact value for $IP$ from it are our results. Theorem \ref{MS} is a slightly improved version of the one in the book, this version is from \cite{L89}.

\subsection{Fooling Sets}
Fooling sets are implicitly used in \cite{Y79} by Yao and defined in \cite{LS81} by Lipton and Sedgewick.

\begin{defi} The elements of a \emph{1-fooling set} $S_1$ are input-pairs $(x^{(i)},y^{(i)})$ with the following two properties: $\forall i \;\; f(x^{(i)},y^{(i)}) = 1$ and $\forall i \neq j \;\; f(x^{(i)},y^{(j)}) = 0$ or $f(x^{(j)},y^{(i)}) = 0$.\\
Similarly, $S_0$ is a \emph{0-fooling set} if $\forall i \;\; f(x^{(i)},y^{(i)}) = 0$ and $\forall i \neq j \;\; f(x^{(i)},y^{(j)}) = 1$ or $f(x^{(j)},y^{(i)}) = 1$.\\
We call a set $S$ a \emph{fooling set} if it is a 1-fooling set or a 0-fooling set.
\end{defi}

\begin{claim} $C_1(f) \geq |S_1|$ if $S_1$ is a 1-fooling set for $f$.
\end{claim}
\begin{proof}
It is enough to show that each element of a fooling set is in a different 1-rectangle. Indeed, if two elements were in the same rectangle, it would not be monochromatic.
\end{proof}

\begin{cor} $C_0(f) \geq |S_0|$ if $S_0$ is a 0-fooling set for $f$.
\end{cor}

Combining this claim with Claim \ref{DC1} we get:


\begin{cor} $D(f) \geq 1 + \log |S|$ if $S$ is a fooling set for $f$.
\end{cor}

With the help of this fact, we can prove that our upper bounds were tight for $EQ$, $NE$, $GT$ and $DISJ$.

\begin{claim} $D(EQ) = n + 1$.
\end{claim}
\begin{proof}
$S_1 = \{(x,x): x \in \{0,1\}^n\}$ is a 1-fooling set. Thus $D(EQ) \geq 1 + \log (2^n) = n + 1$.
\end{proof}

\begin{cor} $D(NE) = n + 1$.
\end{cor}

The same set is a fooling set for $GT$ as well.

\begin{cor} $D(GT) = n + 1$.
\end{cor}

\begin{claim} $D(DISJ) = n + 1$.
\end{claim}
\begin{proof}
$S_1 = \{(\underline{x},\underline{1}-\underline{x}): \underline{x} \in \{0,1\}^n\}$ is a 1-fooling set. Thus $D(DISJ) \geq n + 1$.
\end{proof}

However, for $IP$ we cannot find a sufficiently large fooling set. (We could not prove that there are not any but we think so.) For this, we need different lower bound techniques.

\subsection{Rank Lower Bound}
This lower bound was discovered by Mehlhorn and Schmidt \cite{MS82}. We denote the matrix associated to $f$ by $M_f$, the \emph{rank} of the matrix $M$ by $r(M)$. The proof of their theorem is based on the following simple fact well-known from linear algebra:

\begin{claim} $M = A + B \Rightarrow r(M) \leq r(A) + r(B)$.
\end{claim}

We can imagine each step of the protocol as cutting the matrix $M$ into two smaller matrices. We denote these by $M_0$ and $M_1$.

\begin{cor} $r(M) \leq r(M_0) + r(M_1)$
\end{cor}
\begin{proof}
If we add 0s to $M_0$ and $M_1$ to make them as big as $M$ was, we get two matrices, $A$ and $B$, for which $r(A) = r(M_0)$ and $r(B) = r(M_1)$ and $A + B = M$. Now we can use the previous claim and we are done.
\end{proof}

So in each step of the protocol the rank of one of the remaining matrices is at least half the rank of the previous matrix. Furthermore, we can assume that it has at least one 0 entry (unless $f$ is constant 1 but we can exclude this case). At the end, we get a 0-monochromatic matrix whose rank is $0$. Therefore we derived:

\begin{thm} \label{MS} $D(f) \geq 1 + \log r(M_f)$ unless $f$ is constant.
\end{thm}

With the help of the theorem, we can give a lower bound for $D(IP)$.\\ $M_{IP} \cdot M_{IP} = M^2_{IP}$ is a very simple matrix; the first row and column are all $0$s, the diagonal of the rest is filled with $2^{n-1}$ and all the other entries are $2^{n-2}$. Its rank is $2^n - 1$. From the well-known linear algebraic fact that $r(AB) \leq r(A)$, it follows that $r(M_{IP}) \geq 2^n - 1$. From the theorem we got $D(IP) \geq 1 + n$. This is exactly what we wanted.\\

It is a major open question, also known as the \emph{log rank conjecture} whether $D(f) \stackrel{?}{=} (\log r(M_f))^{O(1)}$, whether the complexity can be bounded from above by a polinom of $\log r(M_f)$. The largest gap has been showed by Nisan And Wigderson in \cite{NW94}. They exhibit a function for which $D(f) = \Omega(n)$ and $\log r(M_f) = O(n^{1/\log 3}) = O(n^{0.631 \ldots})$.

\subsection{Discrepancy}
This lower bound has a parameter, a probability distribution $\mu$ over the elements of $M_f$. If $\mu$ is the uniform distribution, it equals to the number of elements in a given set divided by $2^{2n}$. Let us denote the maximum of the measures of monochromatic rectangles by $w$.

\begin{claim} $C^D(f) \geq 1/w$
\end{claim}
\begin{proof}
Each leaf can have measure at most $w$, the leaves partition the matrix, the matrix has measure $1$, we are done.
\end{proof}

If the measure is concentrated only to the 1 (or 0) entries of the matrix, we denote it by $\mu_1$ (or $\mu_0$). The maximum of the measures of 1-rectangle's (or 0-rectangle's) is denoted by $w_1$ (or $w_0$). With the same prove as before we get the following lower bounds:

\begin{claim} \label{rlb} $C^D_0(f) \geq 1/w_0$ and $C^D_1(f) \geq 1/w_1$
\end{claim}

Now let us examine how we can apply these bounds on $IP$.

\begin{claim} If $R \subseteq M_{IP}$ is a 0-rectangle, then $|R|$ (the size of $R$) is at most $2^n$.
\end{claim}
\begin{proof}
We will use linear algebra. We imagine the inputs as vectors over $GF(2)$. We denote the set of rows (or columns) of $R$ by $A$ (or $B$). Because $R$ is a 0-rectangle, we have $A \bot B$. This implies that $\dim \left\langle A \right\rangle + \: \dim \left\langle B \right\rangle$ (the sum of the dimensions of their generated subspaces) is at most $n$. Therefore $|R| = |A| \cdot |B| \leq 2^n$.
\end{proof}

\begin{claim} \label{IP1} If $R \subseteq M_{IP}$ is a 1-rectangle, then $|R|$ (the size of $R$) is at most $2^{n-1}$.
\end{claim}
\begin{proof}
Like in the previous proof, we are using linear algebra again. We put an extra 1 digit to the end of each vector to make them pairwise perpendicular. We denote these new sets of vectors by $A'$ and $B'$. Fix an arbitrary element $a \in A'$. Note that $(a + A') \cap A' = \emptyset$ because $a + A'$ always ends with a 0. Therefore $|\left\langle A' \right\rangle| \geq 2|A|$ and similarly $|\left\langle B' \right\rangle| \geq 2|B|$. $\dim \left\langle A' \right\rangle + \: \dim \left\langle B' \right\rangle \: \leq n + 1$ implies $2^{n+1} \geq |\left\langle A' \right\rangle| \cdot |\left\langle B' \right\rangle| \geq 2|A'| \cdot 2|B'| = 4|A| \cdot |B| = 4|R|$.
\end{proof}

From any of the above claims we can prove $D(IP) \geq n + 1$ with the help of the discrepancy lower bound. The number of 0s in $M_{IP}$ is $2^{2n-1} + 2^{n-1}$ while the number of 1s is $2^{2n-1} - 2^{n-1}$. Let us take the uniform measure $\mu_0$ concentrated only to the 0 entries of $M_{IP}$. The largest 0-rectangle's measure is at most $2^n /(2^{2n-1} + 2^{n-1}) < 2^{-n+1}$. Therefore from Claim \ref{rlb} we have $C^D_0(f) > 1/2^{-n+1} = 2^{n-1}$. From Corollary \ref{DC0} we have $D(IP) > n$.\\
In a similar argument we could have used $\mu_1$ and 1-rectangles as well, that gives $C^D_1(f) \geq 2^n - 1$ and the same lower bound. So we have proved again $D(IP) = n + 1$.\\

\clearpage
\section{Random EQ}

\subsection{Basic Definitions}
We distinguish between two types of random protocols; whether the random coin flips are $private$ or $public$. This means whether each of them can generate a random string for oneself only or there is a random string that both of them can see. In this paper we are only dealing with bounded error protocols, for unbounded errors see \cite{PS86}. This part follows the book (\cite{KN96} pp. 28-34) but Claim \ref{Rdeltaepsilon} what is stated in a false form in the book (pp. 30) was corrected. The notations are:

\begin{defi} $ $
\begin{itemize}
    \item For $0 < \epsilon < 1/2$, $R_\epsilon(f)$ is the minimum height of a (private) randomized protocol that computes $f$ with error $\leq \epsilon$. We also denote $R_{1/3}(f)$ by $R(f)$. This is called (two-sided error) probabilistic protocol.
    \item For $0 < \epsilon < 1$, $R_\epsilon^1(f)$ is the minimum height of a (private) randomized protocol that computes $f$ with error $\leq \epsilon$ if $f(x,y) = 1$ and makes no error at all if $f(x,y) = 0$. We also denote $R_{1/2}^1(f)$ by $R^1(f)$. This is called one-sided error probabilistic protocol.
    \item $R_\epsilon^0(f) := R_\epsilon^1(\bar f)$
    \item In case of random public coins, we denote the similar values respectively by $R_\epsilon^{pub}(f)$, $R_\epsilon^{1,pub}(f)$ and $R_\epsilon^{0,pub}(f)$.
\end{itemize}
\end{defi}

If we have a one-sided error protocol, we can get one with smaller error probability but with two-sided error:

\begin{claim} \label{rp-bppE} $R_{\epsilon/(1+\epsilon)}(f) \leq R_\epsilon^1(f)$, \;$R^{pub}_{\epsilon/(1+\epsilon)}(f) \leq R^{1,pub}_\epsilon(f)$
\end{claim}
\begin{proof}
The protocols differ only at the very end; when the last bit (the result) is sent, sometimes instead of a 0 (what might be wrong), they might send a 1. If the output of the one-sided protocol was 0, they will output 0 with probability $\alpha$ in the two-sided case. (So this implies, that the mistake will be $1-\alpha$ if $f(x,y) = 0$.) If $f(x,y)=1$, then they will answer 1 with probability at least $1-\epsilon + \epsilon (1-\alpha) = 1-\epsilon\alpha$, so the chance of making a mistake is $\leq \epsilon\alpha$. If we set $\epsilon\alpha = 1-\alpha$, then we get the optimal result because the chances of making a mistake are equal for both values of $f$. So $\alpha=1/(1+\epsilon)$, this gives $\epsilon/(1+\epsilon)$ chance of error.
\end{proof}

\begin{cor} \label{rp-bpp} $R(f) \leq R^1(f)$, \;$R^{pub}(f) \leq R^{1,pub}(f)$.
\end{cor}

Of course, $R_\epsilon(f)$ and $R^1_\epsilon(f)$ depend on $\epsilon$ only up to a constant factor. It is easier to prove for the one-sided error case, we start with this.

\begin{claim} For $0 < \delta < \epsilon < 1$, \;$R_\delta^1(f) \leq R_\epsilon^1(f) \log \delta / \log \epsilon$.
\end{claim}
\begin{proof}
If we have a protocol for $\epsilon$, all we have to do is to repeat it a few times to get one for $\delta$; if the answer is 1 in any of the cases, we know for sure that $f(x,y) = 1$ indeed. If all the answers are 0, then we answer 0. It is clear that this is a one-sided error protocol. What is the chance that $f(x,y) = 1$ but all our answers are 0 after $t$ repetitions? It is at most $\epsilon^t$. So it is enough to repeat our protocol $\log_\epsilon \delta$ times.
\end{proof}

\begin{claim} \label{Rdeltaepsilon} For $0 < \delta < \epsilon < 1/2$, \;$R_\delta(f) \leq O(\frac{R_\epsilon(f) \log 1/\delta}{(1/2 - \epsilon)^2)})$.
\end{claim}
\begin{proof}
If we have a protocol for $\epsilon$, all we have to do is to repeat it a few times to get one for $\delta$; after repeating it $t$ times, we answer what the majority of the $t$ answers were. (For simplicity, let us suppose that $t$ is odd.) What is the chance that our algorithm is mistaking at least $t/2$ times if we are repeating it $t$ times? Let us denote the indicator of the event that the $i$th answer is wrong by $Z_i$. By Chernoff's inequality, $Pr[(1/t \sum_{i=1}^t Z_i - \epsilon)  \geq x] \leq e^{-x^2 t/2}$. In our case the answer will be the majority of the answers, thus we want to choose $x$ as large as possible, to get the best bound achievable this way, but satisfying that $x + \epsilon \leq 1/2$ (because of this the chance of a wrong answer will be smaller than the right-hand side). This means of course $x = 1/2 - \epsilon$. We have to choose $t$ such that the right-hand side equals $\delta$. This yields $t = -2\log_e \delta /(1/2 - \epsilon)^2$.
\end{proof}

\begin{cor} For $0 < \epsilon < 1/2$, \;$R_\epsilon(f) \leq O(R(f) \log 1/\delta)$.
\end{cor}

A similar statement holds for the public coin model as well. More surprisingly, it is even possible to switch from the private coin model into the public coin model for the cost of $\log n$ bits of communication:

\begin{claim} \label{pub-priv} For $0 < \delta, \epsilon, \delta + \epsilon < 1/2$, \;$R_{\epsilon + \delta}(f) \leq R_\epsilon^{pub}(f) + \log n + 2\log 1/\delta + O(1)$.
\end{claim}
\begin{proof}
The idea is that \A sends the results of her private coinflips to \B so they can simulate public coins. We only have to show that for each public coin protocol we can construct another one using only a few flips without significantly increasing the chance of mistake.\\
For the input-pair $(x,y)$ and (public, unknown length) random string $r$ let $Z(x,y,r)$ denote the indicator of the event that our answer is wrong. For any fixed $x$ and $y$, what is the chance that picking $t$ random strings (out of all possible random strings) at least $(\epsilon + \delta)t$ of our answers are wrong? By Chernoff's inequality: $Pr[(1/t \sum_{i=1}^t Z(x, y, r_i) - \epsilon)  \geq \delta] \leq e^{-\delta^2 t/2}$. If we choose $t = O(n/\delta^2)$, then the right hand side is smaller than $2^{-2n}$. This means that there are $t$ such random strings that for all $x$ and $y$ this probability is smaller than 1, so if we choose instead of all random strings only from this $t$, the chance of making a mistake will be smaller than $\epsilon + \delta$. Sending which random string was chosen requires transferring $\log t = \log n + 2\log 1/\delta + O(1)$ bits, this completes our proof.
\end{proof}

Note that we cannot get rid of this $\log n$ because $R(EQ) = \Theta(\log n)$ while $R^{pub}(EQ) = 1 + 1$. We will show the constructions in the next section, here we include another useful lemma that bounds the gap between random and deterministic communication complexity:

\begin{lem} \label{det-rnd} $R(f) = \Omega(\log D(f))$.
\end{lem}
\begin{proof}
It is sufficient to prove the following: $D(f) \leq 2^{R_\epsilon(f)}(\log 1/(1/2 - \epsilon) + R_\epsilon(f))$. So we have to construct a deterministic protocol from a random one. The basic idea is that for all the possible $2^{R_\epsilon(f)}$ communication-strings \A calculates the chance of the string \emph{from her side}; this means she goes through all the bits of the string and each time when the communication before that given bit implies that it is her turn, she calculates the chance that she would send that given bit and multiplies these probabilities. Then she sends this real number to \Bx . After this \B can calculate, with his probabilities attached to each string, the chance of each possible communication-string. Summing these for the cases that give a 1 answer, he gets either at most $\epsilon$ or a least $1-\epsilon$, so he will know for sure whether the answer is 0 or 1. The only problem is that she cannot send the exact values because they are real numbers, so she has to round them. We have to determine how accurately she should send these real numbers to have a small rounding mistake.\\

It is enough to send $\log 1/(1/2 - \epsilon) + R_\epsilon(f)$ bits of accuracy each time, doing so the rounding mistake at each bit-string is at most $2^{-(\log 1/(1/2 - \epsilon) + R_\epsilon(f))} = (1/2 - \epsilon)/2^{R_\epsilon(f)}$, thus the total rounding mistake for all the $2^{R(f)}$ cases together is less than $1/2 - \epsilon$. If we add this to the original chance of error what was $\epsilon$, it is still less than $1/2$, so \B can decide safely whether the answer is 0 or 1.
\end{proof}

\subsection{Lower Bounds}
In this section we mainly deal with the $EQ$ function but sometimes our lower bounds hold for other functions as well. This section is completely our result. The easiest way to obtain a lower bound for $EQ$ is to simply apply Lemma \ref{det-rnd} using $D(EQ) = n + 1$:

\begin{cor} $R(EQ) = \Omega(\log n)$.
\end{cor}

If we have a closer look at that proof, we can see that in fact there we derived the following lower bound:

\begin{claim} $R(EQ) \geq \log (n + 1) - \log \log (6(n+1))$.
\end{claim}
\begin{proof}
In Lemma \ref{det-rnd} we have proved: $D(f) \leq 2^{R_\epsilon(f)}(\log 1/(1/2 - \epsilon) + R_\epsilon(f))$. This gives in our case: $n + 1 \leq 2^{R(EQ)}(\log 6 + R(EQ))$ thus by Corollary \ref{L-shift} we have $R(EQ) \geq \lambda(6(n+1)) - \log 6$, by Claim \ref{L-est} we are done.
\end{proof}

There is a completely different way to prove a similar lower bound. First we prove it only for the \textit{one-way} case (this means that only \A is speaking until the end, then \B tells the result) because it is easier to understand, then for the general case.

\begin{claim} $R_{1/4}^{one-way}(EQ) \geq \log n$.
\end{claim}
\begin{proof}
Let us suppose that the chance of making a mistake is at most $\epsilon$ and that \A says less than $\log n$ bits. This means that there are only $m < 2^{\log n} = n$ possible bit-strings. For each of them \B has some probability to say 0 or 1. We denote these probabilities by $q_1(y), \ldots , q_m(y)$ if his input is $y$. For fix $y$ each $i$ either $q_i(y) > 1/2$ or $q_i(y) \leq 1/2$. This is $2^m < 2^n$ possibilities. So there are $x,y$ inputs such that for all $i$ both of them is bigger or at most $1/2$, therefore $\left| q_i(x) - q_i(y) \right| \leq 1/2$.\\

Let us suppose that \Ax 's input is $x$. If \Bx 's input is also $x$, he has to say 1 with probability at least $1-\epsilon$ but if his input is $y$, he can say 1 with probability at most $\epsilon$. Let us denote the chance that \A sends the $i$th possible bit-string if her input is $x$ by $p_i$. Then we have:
$$ 1 - \epsilon - \epsilon \leq \sum p_i q_i(x) - \sum p_i q_i(y) = \sum p_i (q_i(x) - q_i(y)) \leq \sum p_i 1/2 = 1/2$$
So $\epsilon > 1/4$ if the communication ends after less then $\log n$ bits. This proof works even if $\epsilon = 1/4$ by examining a few cases however we omit this part of the proof because it is not interesting and our next claim will be stronger.
\end{proof}

\begin{claim} $R(EQ) \geq \log n - \log \log n$.
\end{claim}
\begin{proof}
We will use the well known fact that $(1-1/n)^n \geq 1/4$.\\

Let us suppose that we have a given random, private coin protocol that finishes in $d$ steps and errs with probability at most $\epsilon$. The basic idea is the following: If for a fixed pair of inputs $x, y$, \A sends 0 or 1 with the same probability in each step, then the chance that the communication will differ is small.\\

Let us suppose that we are in the $i$th step of communication. So far $i-1$ bits have been sent, this is $2^{i-1}$ possibilities. If it is \Bx 's turn, \A is not doing anything; if it is \Ax 's turn, then she has a function that tells the chance for each input of sending a 0 or a 1. We only consider whether this chance is between $0$ and $1/t$ or between $1/t$ and $2/t$ etc. where $t$ is a parameter to be fixed later. This is $t^{2^{(i-1)}}$ possibilities. Through the whole communication, even if she is speaking all the time (like usually women do), this gives only $\prod_{i=1}^d t^{2^{i-1}}$ possibilities. If this is smaller then $2^n$, then we have two inputs, $x,y$, that are the same with probability at least $(1-1/t)^d = ((1-1/t)^t)^{d/t} \geq (1/4)^{d/t} =: \delta$. If \Bx 's input is $x$, then the result has to be 1 with probability at least $1 - \epsilon$ if \Ax 's input is also $x$ but it can be 1 with probability at most $\epsilon$ if \Ax 's input is $y$. Therefore $1 - \delta \geq$ Pr[the communication is different for $(x,x)$ and $(y,x)$] $\geq (1 - \epsilon) - \epsilon$, thus $2\epsilon \geq \delta$. If we want to prove for $\epsilon$, we should choose $t$ such that $\delta  = (1/4)^{d/t} = 2\epsilon$. This implies $t = \frac{d}{\log_{1/4} 2\epsilon} = \frac{-2d}{\log 2\epsilon}$.\\

Therefore from $2^n \leq \prod_{i=1}^d t^{2^{(i-1)}} = t^{\sum 2^{(i-1)}} \leq t^{2^d} = (\frac{-2d}{\log 2\epsilon})^{2^d}$, we get a lower bound for $d$ if the error is at most $\epsilon$. In the default $\epsilon = 1/3$ case, we get $2^n \leq (\frac{-2d}{\log 2/3})^{2^d} \leq (2d)^{2^d}$, so $n \leq 2^d \log (2d) \leq \Lambda(d)$, thus we get $d \geq \lambda(n) \geq \log n - \log \log n$. So we have proved: $R(EQ) \geq \log n - \log \log n$, we are done.\\

Note that in fact for any fixed $\epsilon$, we have that $n \leq \Lambda(d)$ if $n$ is big enough, thus $R_{\epsilon}(EQ) \geq \log n - \log \log n$ if $n$ is big enough.
\end{proof}

Note that in the proof we did not use anything about $EQ$ except that all rows are different. So our lower bound holds for all functions, except those that have the same row twice, but we can suppose that this does not happen, otherwise we could simply forget one of the rows.

\begin{thm} $R(f) \geq \log n - \log \log n$ if all the rows of $M_f$ are distinct.
\end{thm}

This gives a slightly better bound than Lemma \ref{det-rnd} and much better if $D(f)\!\!< n$.

\subsection{Upper Bounds}
The most interesting is that although we know several random
protocols that run in $O(\log n)$ time and compute $EQ$ but none
of them is truly constructive; they all either contain a big,
fixed random set or need to find a big random prime. This might be
because in the proof of Claim \ref{pub-priv} we used Chernoff's
inequality and that is why we cannot transform the public coin
protocol that needs only 2 steps into an explicit one. In this
section we give a new algorithm and compare it with the ones
previously known.

\begin{claim} \label{RpubEQ} $R^{pub}(EQ) = 1 + 1$.
\end{claim}
\begin{proof}
Obviously, it cannot be 1, because then the first player should output the answer without the second player speaking anything.
Now we present two constructions that give $R^{0,pub}(EQ) = 1 + 1$. We can turn these one-sided error protocols into two-sided ones using Corollary \ref{rp-bpp}, so we will be done.\\

Partitioning Construction: We need $2^n$ public random bits. We think about them as the characteristic vector of a subset of $\{0,1\}^n$. \A sends to \B whether her input is in the set or not. Now \B computes. If only one of their inputs is in the set, then the answer must be 0. If both of their inputs are in or out of the set, he answers 1. If their inputs were the same indeed, then he answered 1, so they made no mistake. If their inputs were different, then with probability $1/2$, only one of the inputs was an element of the set, so their answer is 0 with probability $1/2$, just what we wanted.\\

Inner Product Construction: Here we need only $n$ public random bits denoted by $z$, we think about them (and also about $x$ and $y$) as a vector over $GF(2)$. \A sends to \B $\left\langle x,z \right\rangle$. \B compares it with $\left\langle y,z \right\rangle$. If they differ, he answers 0, because $x$ and $y$ must be different. If they are the same, he answers 1. We only have to prove that if they differ, then the chance that $\left\langle  x,z \right\rangle$ and $\left\langle y,z \right\rangle$ also differ is 1/2. Fix a bit where $x$ and $y$ differ. With probability $1/2$, this bit of $z$ equals 1, with probability $1/2$, this bit of $z$ equals 0. Changing only on this bit and leaving the other bits unchanged, exactly one of their scalar products change. This completes the proof.
\end{proof}

Combining this result with Claim \ref{pub-priv} and with the lower bound, we get:

\begin{cor} \label{eqlogn} $R(EQ) = \log n + O(1)$
\end{cor}

Comparing this with the lower bounds, this is almost the best that we can get. However, this is not a constructive proof, it only shows that there $exists$ a proper algorithm, but we have no clue how to construct it. In the remaining part of this section, we present some algorithms that are more constructive but give worse bounds.\\

Another way to achieve a similar upper bound, using a random prime, is the following result of Rabin, Simon and Yao (see \cite{L89'}):

\begin{claim} $R^0(EQ) \leq 4 \log n + 2 + 1$.
\end{claim}
\begin{proof}
We pick a $p$ prime at random from $n^2$ to $2n^2$. This is done by \Ax 's random string. She sends to \B $p$ and $x \mod p$. This requires $4 \log n + 2$ bits. \B compares $y \mod p$ and $x \mod p$. If they are the same, he answers 1, otherwise 0.\\

To show that this protocol is correct, we only have to show that there is a good chance that a random $p$ will not divide $x-y$ unless $x=y$. In fact, because of $\left|x-y\right| \leq 2^n$, it can have at most $log_{n^2} 2^n = n/ (2\log n)$ prime divisors that are bigger than $n^2$. And we know from number theory that the number of primes between $n^2$ and $2n^2$ is approximately $2n^2/\log_e (2n^2) - n^2/\log_e (n^2) \geq cn^2/(2\log n)$ where c is a fix constant. Thus the probability that $p$ divides $x-y$ is smaller than $\frac{n/ 2\log n}{cn^2/2\log n} = 1/cn$, so $R^0_{1/cn}(EQ) \leq 4 \log n + 2 + 1$.
\end{proof}

There is another method using a prime number and giving the same result but the difference is that this one needs only one fixed prime (depending of course on $n$). This appeared in the book (\cite{KN96} pp. 30-31). We are going to introduce a parameter to get different upper bounds for different error-tolerance.

\begin{claim} $R^0_{n/m}(EQ) \leq 2 \log m + 2 + 1$.
\end{claim}
\begin{proof}
Pick any prime $p$ from $m$ to $2m$. (The existence of such a prime follows from Chebyshev's theorem.) We represent the inputs as polynomials over $GF(p)$: $A(z)=x_{n-1}z^{n-1} + \ldots x_1z + x_0$ and we similarly obtain $B(z)$. Now \A picks a random number $z_0$ over $GF(p)$. She sends $z_0$ and $A(z_0)$ to \Bx . \B compares $A(z_0)$ and $B(z_0)$. If they are different, then the answer must be 0, otherwise he answers 1. If $A(z_0) = B(z_0)$, but $A(z) \not\equiv B(z)$ then $z_0$ is the root of $A(z)-B(z)$ what is a polynomial with degree $\leq n-1$. The chance that we picked a root at random is less than $n/p \leq n/m$. So we obtained $R^0_{n/m} \leq 2 \log m + 2 + 1$.
\end{proof}

Choosing $m=2n$ we have:

\begin{cor} $R^0(EQ) \leq 2 \log n + 4 + 1$.
\end{cor}

This is worse with a factor of 2 than Corollary \ref{eqlogn}, but it is a bit more constructive; if we have to work with a fixed $n$ (usually this is the case in applications), we can include any $p$ in the algorithm, \A does not have to search a random prime again and again like in the previous algorithm. Another advantage compared to the previous algorithm is that we can quickly repeat the test by sending a new random element from $GF(p)$. For a fix error tolerance $\epsilon$ choosing $m = \left\lceil n/\sqrt[k]{\epsilon} \right\rceil$ and repeating it $k$ times this gives: 

\begin{thm} $R^0_\epsilon(EQ) \leq (k + 1) \log n + \frac{k+1}{k} \log 1/\epsilon + O(1)$ using a constructive algorithm.
\end{thm}

If $\epsilon \geq 1/n$, this bound is the strongest in the case when $k=1$, so for a ``big'' $\epsilon$ this algorithm should not be repeated at all, it is better if we pick a small $m$. In this case we obtain:

\begin{cor} $R^0_\epsilon(EQ) \leq 2\log n + 2\log 1/\epsilon + O(1)$ using a constructive algorithm.
\end{cor}

If $\epsilon = 1/n^d$ then we have to choose $k \approx \sqrt{d}$, this yields approximately $(d + 2\sqrt{d} + 1)\log n = \log 1/\epsilon + 2 \sqrt{\log 1/\epsilon} + \log n$ bits of communication.\\
If $\epsilon$ is even smaller then we can only get a superlogarithmic bound.\\

Now we examine a generalization of the Partitioning Construction that gives a better upper bound for a ``big'' $\epsilon$ error-tolerance. Instead of dividing the set of inputs into $2$ parts, we divide them into $k$ parts and combine it with Claim \ref{pub-priv}.\\

Pr[$x$ and $y$ are in the same partition] $= 1/k$. To send which partition an input belongs to, requires $\log k + 1$ bits. Thus $R^{0,pub}_{1/k}(EQ) \leq \log k$. If we use Claim \ref{pub-priv}, we get $R^0_{\delta + 1/k}(EQ) \leq \log k + \log n + 2\log 1/\delta + O(1)$. How to choose $k$ and $\delta$ for a given $\epsilon$ error-tolerance? We have $\delta + 1/k = \epsilon$, and our goal is to minimize $\log k + 2\log 1/\delta$. We can simply do this by transforming and applying the arithmetic-qubic mean inequality: $\log k + 2\log 1/\delta = -\log (\frac{1}{k \delta^2}) = 1 -\log (\frac{2}{k \delta^2}) \geq 1 -\log ((\frac{2/k + \delta + \delta}{3})^3) = 1 -3\log (\frac{2}{3 \epsilon})$ and the equality holds iff $\delta = 2/k$. So $\epsilon = 3/k$,  and for the overall complexity is $\log n + 3\log k + O(1)$. If we have an $\epsilon$, we can pick $k = \left\lceil 3/\epsilon \right\rceil$. This gives the following upper bound:

\begin{thm} $R^0_\epsilon(EQ) \leq \log n + 3\log 1/\epsilon + O(1)$
\end{thm}

This is slightly better than the one we got by applying the prime-method if $\epsilon \geq 1/n$ and does not need any prime testing or number theory at all. It is less constructive in the sense that we have no idea at all how to find suitable random strings that we got from Chernoff's inequality. But if we are working with a fix $n$, then we can include the necessary random strings, therefore this is the best of all the above algorithms if $\epsilon$ is bigger than $1/n$ and this seems to be realistic in most applications.

\clearpage
\section{The Direct-sum problem}

\subsection{The Problem}

The Direct-sum problem arises not only in CC but in almost all computational models and is yet unsolved basically in all of them. It was introduced to CC by Karchmer et. al \cite{KRW91}. The problem is simply this question: Can it be easier to solve two independent problems at the same time than solving them one after the other?\\

In CC, the problem has several versions. To state them in a nice form, we introduce first some notations. If we have two functions, $f$ and $g$, and both \A and \B have two inputs: $x_f$, $x_g$ and $y_f$, $y_g$ and they would like to compute both $f(x_f,y_f)$ and $g(x_g,y_g)$ then we denote this problem by $f \times g$. In the case where they would like to know only $f(x_f,y_f)\wedge g(x_g,y_g)$ we denote it by $f \wedge g$. The problem is also interesting when $f$ and $g$ are the same; in this case they both hold $k$ inputs (($x_1, \ldots , x_k)$ and ($y_1, \ldots , y_k)$) and would like to compute $f(x_i, y_i)$ for every $i$. We denote this version by $\times _k f$. In the case where we want to know only whether all the outputs are 1, we denote it by $\wedge _k f$.\\

Now we can state the following conjectures:

\begin{conj} \label{D-s}
\item[(i)] $D(f \wedge g) = D(f) + D(g)$
\item[(ii)] $D(f \times g) = D(f) + D(g)$
\item[(iii)] $D(\wedge _k f) = k \cdot D(f)$
\item[(iv)] $D(\times _k f) = k \cdot D(f)$
\end{conj}

Obviously the right-hand side is always bigger than the left.\\
It is clear that (i) is weaker than (ii) and (iii) is weaker than (iv), moreover (i) is weaker than (iii) and (ii) is weaker than (iv). So (iv) is the strongest and (i) is the weakest. Although we can not prove for general functions any of them, we know that they hold for many specific functions, like $EQ$, $GT$ or $DISJ$. Another useful definition of the topic is $AMT(f) = \lim _{k \rightarrow \infty} D(\times _k f)/k$, it is called \textit{Amortized Time Complexity}.

\subsection{Constant Factor Difference in the Direct-sum problem}
Here we give a few counterexamples for the current form of Conjecture \ref{D-s} and then modify it to a form for which we cannot present any counterexamples.\\

If $f$ is EQ for $N = 5$ (they both hold a number from 1 to 5), then $D(f \times f) = 5 + 2 \neq 2 \cdot (3 + 1) = 2 \cdot D(f)$. This counterexample works because the information is sent in bits, we can not get rid of Base 2. If we were allowed to send instead of bits any arbitrary amount of information (but of course the amount of information should be specified in the protocol), we could correct this mistake. The cost of one step would be the logarithm of the information send. We denote this complexity by $\tilde{D}(f)$.\\

Eg., in the previous case, \A could send her number first (costing $\log 5$) and then \B could send her back the result, this implies $\tilde{D}(f) \leq \log 5 + 1$. In the Direct-sum case, the cost would be $\log (5 \cdot 5) + 2$ = $2 \cdot \tilde{D}(f)$, so our counterexample does not work for this little modification.\\

However, this is not the only type of counterexample that we know. Consider the following example: Let \Ax 's input be $2 + 4$ bits and \Bx 's be $1 + 4 + 16$ bits. The first 2 bits of \A are pointing to a bit in the 4-bit block of \Bx , the other 4 are pointing to a bit from the 16-bit block of \Bx . If the first bit of \B is 0, then the value of the function is the bit from the 4-bit block, if it is 1, then the output is the bit from the 16-bit block. We shall denote this $f$ for later reference by $TAB$($2-4$) ($There And Back$ $2-4$). It is easy to see that $D(f) = 5 + 1$ and we cannot improve significantly even if we do not insist on the bit-wise communication but allow an arbitrary amount of information in each step.\\

But we have a solution for $f \times f$ in $9 + 2$ steps:

\begin{itemize}
\item[-] If both of the first bits of the inputs of \B are 1, then he sends a $0$, then \A replies with $2$ times $4$ bits, finally he sends back the result ($2$ bits), this altogether yields $11$.
\item[-] If one of his first bits is 1 while the other is 0, then he sends $10$, then a $0$ or $1$ depending which input contains the $1$. She replies with $4 + 2$ bits and he sends back the result ($2$ bits), this altogether yields $11$.
\item[-] If both of the first bits of the inputs of \B are 1, then he sends $11$. She can reply with $2 + 2$ bits, he sends back the result ($2$ bits), this altogether yields only $8$.
\end{itemize}

Note that with counterexamples of these kinds one may achieve only a constant factor difference in complexity, so if in Conjecture \ref{D-s} we would write an $\Omega$ before the right-hand side, they would not disprove the conjecture. To avoid these counterexamples and have a conjecture that does not need $\Omega$s, we are going to introduce a new model that differs from the classical one only in a constant factor and has no counterexamples at all for Conjecture \ref{D-s}. However, we are only going to deal with the $\times$ case (the (ii) and (iv) parts of the conjecture).

\subsection{Fluent Communication}
Here we introduce a new model of communication that is slightly different from the classical one. We are going to refer to the classical one as \textit{Bit-wise} and to the new one as \textit{Fluent}. The basic idea is that someone might send a piece of information faster than 1, but for the following sacrifice; if the information is not the one she wanted to send fast, she has to send it slower then 1. First, we are going to give a simple example how it works, then it will be followed by the exact definition.\\

Eg., in the first step of a protocol \A would like to send a single bit to \Bx . If the bit is 0, then the function is computed and the communication is over but if it is 1, they have to go on. Now in the case the bit is 0, she has time, but if it is 1, she should hurry to reduce the length of the worst case. We allow her to pick two numbers, $a$ and $b$, such that $1/a + 1/b = 1$ and she can send the 1 in $\log b$ while the 0 should be sent in $\log a$ time. If $b$ is smaller then $2$, they can finish the communication earlier.

\begin{defi} The \textit{Fluent Communication} is very similar to the classical one. But in each step instead of sending a single bit, the player can do the following: First, she has to pick an $n \geq 2$ natural number (she wants to send an information of size $n$) and $a_1, \ldots , a_n$ reals such that $1/a_1 + \ldots + 1/a_n = 1$. She can send the first type of information in $\log a_1$ time, the second in $\log a_2$ time etc. (She cannot pick the numbers in the middle of the algorithm but she has to pick them before they start to compute the function, so the numbers are built into the protocol.)
\end{defi}

Note that if she could pick the numbers during the algorithm, then she could send information by what numbers she is choosing, we do not want to allow this. If $n = 2$ and $a_1 = a_2 = 1/2$ in each step, then we get the \textit{Bit-wise} communication.\\

In fact, we can assume that $n=2$ in each step. We present a way how to make $n$ become $n - 1$ (if $n \geq 3$) without increasing the time: If in a given step the numbers are $a_1, \ldots , a_n$, then instead of them we can choose the numbers $a_1$ and $\frac{1}{1/a_2 + \ldots + 1/a_n}$ and in the next step $n-1$ numbers: $a_2 (1/a_2 + \ldots + 1/a_n), \ldots , a_n (1/a_2 + \ldots + 1/a_n)$. The sum of the reciprocals is indeed $1$ and $\log \frac{1}{1/a_2 + \ldots + 1/a_n} + \log (a_i (1/a_2 + \ldots + 1/a_n)) = \log a_i$, hence we are done.

\begin{cor} \label{fb} The \textit{Fluent} complexity is always at most the \textit{Bit-wise} complexity.
\end{cor}

Another thing to assume is that \A and \B are switching among each other until the protocol is finished. This is exactly the reverse version of the previous method, we can increase $n$ to $n + 1$ if it is \Ax 's turn again after one of the $n$ possibilities, the proof goes in exactly the same way. Of course we can assume only either this or the previous version.\\

Now we are going to show how the fluent algorithm works for $TAB$($2-4$). We have to use the ``Fluentness'' only at the beginning. If the first bit of \B is 0, he sends it in $\log 5$ time while he sends it in $\log 5/4$ time if it is a 1. (Since $1/5 + 4/5 = 1$, this is a correct step.) In the 0-case, they need $2 + 1$ more bits of communication, in the 1-case $4 + 1$ more, both yielding a total $4 + \log 5/4$ time, better than the \textit{Bit-wise}. Moreover, $2 \cdot (4 + \log 5/4) \leq 11$, the number of bits needed to solve two copies of $TAB$($2-4$), so it can even enhance the solution of two copies.

\begin{claim} The time needed for \textit{Fluent Communication} equals to $\log C^p(f)$.
\end{claim}
\begin{proof} If we have a protocol-tree, we can easily construct a \textit{Fluent} protocol. At each node, we count the number of leaves under its two sons. If they are $a$ and $b$, our two numbers shall be $(a+b)/a$ and $(a+b)/b$. This way the time of the total communication by induction on the number of leaves is $\log a + \log ((a+b)/a) = \log b + \log ((a+b)/b) = \log (a+b)$, just what we wanted to prove.\\

Showing that a fast \textit{Fluent} protocol gives a protocol-tree with few leaves is a similar argument as the one before. Now the induction goes by the number of steps in the protocol. Let us suppose that the protocol finishes in $\log r$ steps in the worst case. Let the numbers of the first step be $a_1, \ldots , a_n$, the number of the leaves in the remaining part of the protocol be $L_1, \ldots , L_n$, respectively. The induction gives $\forall i$ $\log L_i + \log a_i \leq \log r$. After transforming $1/a_i \geq L_i/r$. Summing up and using $\sum 1/a_i = 1$ we get: $1 \geq \sum L_i/r$, thus $r \geq \sum L_i =$ the number of leaves of the protocol-tree we constructed, so just what we wanted to prove.
\end{proof}

Combining with Claim \ref{CpD} we have proved again Corollary \ref{fb}.\\

By the nature of the \textit{Fluent Communication}, none of our old tricks for disproving Conjecture \ref{D-s} works. We can finally restate the conjecture in the desired form:

\begin{conj} $C^p(\times _k f) = k \cdot C^p(f)$
\end{conj}

Note that $C^p(\times _k f) \leq k \cdot C^p(f)$ is trivial. Now we are going to prove that the \textit{Amortized Time Complexity} of $f$ can be at most the \textit{Fluent} computation of $f$:

\begin{thm}
$AMT(f) = \lim _{k \rightarrow \infty} D(\times _k f)/k \leq \log C^p(f)$
\end{thm}
\begin{proof}
We are proving by induction on $C^p(f)$. We assume that $D(\times _k f') \leq g_l(k) \cdot \log l$ for all $l < C^p(f)$ where $g_l(k) \leq k + c_l$ and $c_l$ is a constant depending only on $l = C^p(f')$, the number of leaves of $f'$. If we prove that this holds for $l = C^p(f)$ as well, we have proved the theorem.\\

After sending the first bit, we have two easier problems to solve, let us denote them by $f_m$ and $f_n$. We shall denote $C^p(f)$ by $L$, $C^p(f_m)$ by $M$ and $C^p(f_n)$ by $N$. Note that obviously $M$ and $N$ are both less than $L$, so we can use the induction. We would like to prove the following lemma for a suitable $g_L(k)$:

\begin{lem}
$2^{\left\lfloor g_L(k) \log (M + N)\right\rfloor} \geq \sum {k \choose i} 2^{\left\lceil g_M(i) \log M\right\rceil + \left\lceil g_N(k-i) \log N\right\rceil}$
\end{lem}
\begin{proof}
The proof only consists of some calculation. First we replace the lower and upper integer-parts by -1 and +1: $2^{g_L(k) \log (M + N) - 1} = \sum {k \choose i} 2^{ g_M(i) \log M + 1 + g_N(k-i) \log N + 1}$. Now we get rid of the base two: $1/2$ $(a+b)^{g_L(k)} = 4 \sum {k \choose i} M^{g_M(i)} N^{g_N(k-i)}$. The right side is bounded by induction by $4 \sum {k \choose i} M^{i + c_M} N^{k - i + c_N} \leq 4(M+N)^k M^{c_M} N^{c_N}$. Thus we need: $1/2$ $(a+b)^{g_L(k)} \geq 4(M+N)^k M^{c_M} N^{c_N}$ to hold for all $M + N = L$. If we choose $g_L(k) = \min_{M+N=L} (k + c_N + c_M + \log 8)$, our inequality follows.
\end{proof}

To give a protocol for $\times_k f$, we are going to use the ones for $\times_k f_m$ and $\times_k f_n$. Wlog. we can suppose that $M \leq N$, so $f_n$ is the tougher problem. In the case, when we are going to have $i$ out of the $k$ inputs going toward $m$ and $k-i$ toward $n$, the first half of the message will contain the information which message is going which way, we call these messages respectively $n$- and $m$-cases. The second half will be exactly of length $i g_M(i) \log M + (k-i) g_N(k-i) \log N$ giving enough space to communicate the problems one by one by induction, we have no problem how to do this part. In the first part, we send $g_L(k) \log (M + N) - k g_N(k) \log N$ 0s if all $k$ problems are $n$-cases. So the first half of the message will be something like this: 0000. If we have only $1$ $m$-case and $k-1$ $n$-cases, then we use the first $g_L(k) \log (M + N) - g_M(1) - (k-1) g_N(k) \log N$ bits to specify where the $m$-case is. So if it is the first one, our message should be something like this: 000100 (we have to skip 000000 because all messages starting with 0000 are reserved for the full $n$-case). If it is the second one, something like this: 000101 etc. It is guaranteed that we do not run out of space by the previous lemma; if we run out, then the sum on the right-hand side until that $i$ where we run out of space would exceed the left-hand side yielding a contradiction.\\
We leave the exact details of the proof to the reader.
\end{proof}

\subsection{Communication with Partial Information}
It is possibly to define everything in communication complexity if we allow the players to have a partial information at the beginning of the game depending on their input. In this case the input-pair $(x,y) \in S \subseteq X \times Y$. Surprisingly, we know that the Direct-sum conjecture does not hold in this case. This model was studied by Orlitsky but we follow the book (\cite{KN96} pp. 63-66) that follows the paper by Feder et al. \cite{FKNN91}.\\

We can imagine a partial communication game as a matrix filled
with 0s, 1s and $*$s. The $*$s denote that $(x,y) \notin S$, the
player cannot have this input-pair. Let us denote such a problem
by $f_*$. If we fill in all the starred places by 0s and 1s, we
get a particular classical problem. We denote the set of these
problems by ${\cal F}$.

\begin{claim} \label{nostar} $D(f_*) = min_{f \in {\cal F}} D(f)$.
\end{claim}
\begin{proof}
If we have a protocol for any $f \in {\cal F}$, then it is a protocol for $f_*$ as well.\\
On the other hand, if we have an optimal protocol for $f_*$, then
it gives a rectangle-partition of the whole matrix $X \times Y$
where all rectangles are monochromatic, meaning they contain
either only 0s and $*$s or 1s and $*$s. If we fill in the $*$
entries by 0s and 1s regarding which rectangle they belong to, we
get an $f \in {\cal F}$ for which the same protocol works.
\end{proof}

Now for a while, instead of examining the common 0-1 range case, we will allow functions whose range is different, namely $f(x,y) = x$. While $D(f) = \log \left|X \right|$ and is of no interest in the standard model, we can examine it in the case of partial information where the complexity will depend on $S$. We denote the complexity of such a problem by $D(S)$.\\

Let us consider the following problem: $X = \{0,1 \}^n$, $Y = {X \choose 2}$, thus \A has one number, while \B has two. We are going to denote \Bx 's input $y$ by $[u,v]$. $S = \{x,[u,v] : x=u \; or \; x=v\}$, thus he knows two numbers and she knows one of them and he has to find out which. The name of this problem is $NBA$ because you can think about the inputs as the names of teams. \B is really interested in basketball and knows which two teams played last night and would like to know who won. \A heard the winner in the news but she does not know which two teams played yesterday. How much information is needed to be exchanged?\\

A great advantage of this problem is that we can also think about it as a classical 0-1 range problem; knowing whether $u$ or $v$ equals $x$ is equivalent for \B to know whether the first or second of his teams is the winner. Now let us examine the complexity.\\

If only she is allowed to speak, she has to send $n$ bits, otherwise there would be two unseparated elements of $X$. We denote this by $D^{one-way}(S) = n$. But if both of them can speak, they can be faster.

\begin{claim} \label{NBA} $D(NBA) \leq \log n + 1$.
\end{claim}
\begin{proof}
First \B sends \A an index $i$ for which $u_i \neq v_i$. This
requires $\log n$ bits. Now she can send back $x_i$, this reveals
for him which is the winning team.
\end{proof}

Can this problem be solved faster? The answer is no and it follows
from the following Claim:

\begin{claim} $D(S) \geq \log D^{one-way}(S)$.
\end{claim}
\begin{proof}
Suppose that their is protocol with $D(S)$ steps. We construct
from it a one-way protocol with $2^{D(S)}$ steps. For each
possible $2^D(S)$ bit-strings, \A sends whether they are possible
or not. (Whether her input intersects the leaf belonging to that
bit-string or not.) From this and $y$, \B can determine the only
possible bit-string. This reveals the answer because the original
protocol was good.
\end{proof}

Let us denote the case when first \B is allowed to speak, but after he finishes, only \A can speak. We denote this by $D^2(S)$ and call it a $two-round$ communication problem. As we have seen in the proof of Claim \ref{NBA}, $D^2(NBA)= \log n +1$. Now we are going to prove an interesting theorem, that is $D^2(S) = O(D(S))$. For the proof, we need some preparation.\\

For a problem $S$, we define a hypergraph $G_S = (X,E)$ as
follows: For every $y \in Y$ there is a hyperedge $e_y = \{x :
(x,y) \in S \}$. A coloring of $G_S$ with $c$ colors is a function
$\psi : X \rightarrow \{1, \ldots , c \}$ such that for every
hyperedge its vertices have all different colors. The minimal
number of colors needed is the chromatic number of $G_S$, we
denote it by $\chi (G_S)$. The size of the largest hyperedge is
the degree of $G_S$, we denote it by $d(G_S)$. It is not hard to
see that

$$\left\lceil \log \chi (G_S) \right\rceil = D^{one-way}(S) \geq
D^2(S) \geq D(S) \geq \log (d(G_S))$$.

Now we need a technical claim, that states a suitable family of $hash$-functions exists.

\begin{lem} Let $m$ and $t$ be two arbitrary integers. There are constants $C$ and $\delta$ such that there exists a family $H_{m,t}$ that contains $k= \delta t \log m$ functions whose domain is $\{1, \ldots , m \}$ and whose range is $\{1, \ldots , p= Ct^2 \}$ such that for every $A \subseteq \{1, \ldots , m \}$ of size at most $t$, at least half of the functions from $H_{m,t}$ are injective over $A$.
\end{lem}
\begin{proof}
This is a technical proof using a probabilistic argument. We
choose $k$ functions, $h_1, \ldots , h_k$, at random whose domain
is $\{1, \ldots , m \}$ and whose range is $\{1, \ldots , p \}$
where $C$ is going to be fixed soon. For a fix $A \subseteq \{1,
\ldots , m \}$ of size at most $t$, the probability that a random
function is injective is at least $1 \cdot \frac{p-1}{p} \ldots
\frac{p-t+1}{p} \geq (1-\frac{t}{p})^t = (1-\frac{1}{Ct})^t \geq
3/4$ if $C$ is big enough ($C = O(1)$, so it does not depend on
$t$). Let $Z_i$ denote the indicator of the event that $h_i$ is
injective over $A$. We know that $Pr[Z_i=1] \geq 3/4$. Using
Chernoff's inequality, we obtain $Pr[(1/k \sum_{i=1}^k Z_i) - 3/4
\geq -x] \leq e^{-x^2 k/2}$. If $x=1/4$, the right-hand side is
$e^{-\Theta(k)}$, this is the chance that half of the functions is
injective over $A$. There are $\leq m^t$ possible $t$ element
subsets of $\{1, \ldots , m \}$. So if $m^t \cdot e^{-\Theta(k)} <
1$, then there exists a suitable family of functions and we can
achieve this by choosing $\delta$ small enough, depending on $m$
and $t$.
\end{proof}

\begin{thm} $D^2(S)=O(D(S))$.
\end{thm}
\begin{proof}
We construct a two-round protocol using $G_S$. We fix a coloring $\psi$ of $G_S$ with $\chi(G_S)$ colors and fix a family of functions $H=H_{\chi(G_S),d(G_S)}$ satisfying the conditions of the previous Lemma. The colors used to color the vertices of the edge $e_y$ determine a subset $A \subseteq \{1, \ldots , \chi(G_S) \}$ that has size at most $d(G_S)$, thus it satisfies the condition of the lemma, therefore there is a function $h \in H$ that is injective over $A$.\\
The protocol is simple. \A sends the name of $h$, then \B sends back $h(\psi(x))$ and they are done, he knows $x$ because of the injectivity.\\

Sending $h$ requires $\log \left|H \right| = \log (\delta d(G_S)
\log \chi(G_S) ) = \log \delta + \log d(G_S) + \log \log
\chi(G_S)$ bits. Sending $h(\psi(v_x))$ required $\log (C
(d(G_S))^2) = \log C + 2 \log d(G_S)$ bits. This is together $O(1)
+ O(\log d(G_S)) + \log \log \chi(G_S)$. But we know that $\log
d(G_S) \leq D(S)$ and also $\left\lceil \log \chi(G_S)
\right\rceil = D^{one-way}(S)$ and $\log D^{one-way}(S) \leq
D(S)$. Therefore this is in fact $O(D(S))$ bits of communication.
\end{proof}

Note that we can see from the proof that we have not used that half of the functions from $H$ are injective. This suggests that if they have to solve several problems at the same time, they might be faster because \B might find a function that is injective for many of his inputs at the same time. This is in fact true.

\begin{thm} $D^2(\times_k S) = O(k \log d(G_S)) + \log k \cdot \log \log \chi(G_S))$.
\end{thm}
\begin{proof}
The protocol is almost the same as in the previous theorem. When choosing a hash-function, \B can choose an $h_1 \in H$ such that it is injective for at least half of his inputs. Then he chooses an $h_2 \in H$ injective for at least half of his remaining inputs etc. He sends the names of this $\log k$ functions, then he sends which inputs belong to which functions. Finally \A sends back every $h_{j(i)}(\psi(x_i))$ where $j(i)$ is the index of the proper hash function for each $x_i$ input.\\

Sending the names of the functions requires $O(\log k (\log d(G_S)
+ \log \log \chi(G_S))$ bits. Sending which input belongs to which
function requires $k \log \log k$ bits, but we can use a better
prefix coding; if an input $x_i$ belongs to the $j$th function, he
sends $j-1$ 1s followed by a 0. Because at most $1/2^j$ inputs
belong to $h_j$, this is at most $\sum_1^{\log k} jk/2^j = O(k)$
bits. Sending back the hash-values takes $O(k \log d(G_S))$ bits.
Alltogether this is what we wanted to prove.
\end{proof}

Now let us consider the case when $S=NBA$. We know that this is equivalent to a 0-1 range problem with partial information. It is easy to see that $\chi (G_{NBA}) = 2^n$ and $d(G_{NBA}) = 2$. Hence $D(\times_k NBA) = O(k + \log k \cdot \log n) <\!\!< k \log n = k D(NBA)$. Therefore the Direct-sum conjecture is not true in the partial information case.\\

Note that this does not give us a counterexample in the standard case using Claim \ref{nostar}, because we can fill in the *s with 0s and 1s in the direct-sum version that it does not become the direct sum of any standard function, so the classical conjecture still remains open.\\

\subsection{The Direct-sum problem in the Randomized case}
We can ask the same question in randomized case. The best way to ask is the following: Is it true that $R^{pub}(\times _k f) = k \cdot R^{pub}(f)$? (Again, we are not interested in constant factor difference.) Here, not even $R^{pub}(\times _k f) \leq k \cdot R^{pub}(f)$ is trivial because the chance of mistake commulates. We can only state $R^{pub}(\times _k f) \leq k \cdot R_{1/3k}^{pub}(f) \leq O(k\log k \cdot R^{pub}(f))$ using Claim \ref{Rdeltaepsilon}. In this section we are going to prove that $R^{pub}(\times _k f) = O(k \cdot R^{pub}(f))$ for $f = EQ$. This section is completely our result.\\

\begin{thm} $R^{pub}(\times _k EQ) = O(k)$.
\end{thm}
\begin{proof}
First note that it is sufficient to prove $R_\epsilon^{pub}(\times _k EQ) = O(k)$ for any $\epsilon$ because of Claim \ref{Rdeltaepsilon}. Another useful thing, that it is enough if we present an algorithm with expected running time $O(k)$. Then using the Markov inequality we can get rid of cases when it is running for more then $O(ck)$ increasing the error by at most $1/c$.\\

Claim \ref{RpubEQ} gives $R_\epsilon^{0,pub}(EQ) = 1 + 1$. Moreover, we know that if $x \neq y$, then the chance that the answer is wrong is exactly 1/2. We denote the bit sent by \A by $TEST(x)$ and the bit for which \B replies 1 (accept) by $TEST(y)$. Note that of course when that $TEST$ uses a different random string for each $x_i$. Also note that when they test again for the same $x_i$, $TEST(x_i)$ denotes a different bit. The reader may always easily figure out when a new test is applied. (Usually in the same paragraph $TEST$ denotes the same thing.)\\

First \A sends $TEST(x_i)$ for each $i$. \B compares these with each proper $TEST(y_i)$ and sends back which of them are equal. This takes $O(k)$ bits of communication.\\

For some pairs it turns out that they are not equal, we can forget about them. We group the rest into $pairs$ and for them, she sends $TEST(x_i) \oplus TEST(x_j)$. ('$\oplus$' denotes the sum mod $2$.) He compares this with $TEST(y_i) \oplus TEST(y_j)$ and sends back whether they are equal or not for all pairs. This takes $O(k/2)$ bits of communication.\\

Again, if they are not the same, we know for sure that either $x_i \neq y_i$ or $x_j \neq y_j$. Moreover, the chance that he detects this is exactly 1/2. If $TEST(x_i) \oplus TEST(x_j) \neq TEST(y_i) \oplus TEST(y_j)$, then she sends the same $TEST(x_i)$ (without adding $TEST(x_j)$ to it). If this does not equal $TEST(y_i)$, then we know for sure that $x_i \neq y_i$. If they equal, then $TEST(x_j) \neq TEST(y_j)$, thus $x_j \neq y_j$. We call this a \textit{track-back}. In both cases we have excluded a wrong input-pair and we perform a $TEST$ for the other pair; if the $TEST$ detects a difference again, we know that the other was wrong as well, if it satisfies the $TEST$, we can be more certain that they equal. Discovering each wrong input-pair requires $O(1)$ bits of communication, and after this second step, we have tested each remaining input-pair at least twice.\\

The algorithm goes on like this; we make $four$s, $eight$s and so on until all the $k$ inputs are tested together. (We can suppose that $k$ is a power of $2$.) This requires $O(k) + O(k/2) + O(k/4) + \ldots + O(1) = O(k)$ bits of communication without counting the track-backs at detecting a wrong pair. A track-back at the $l$th round requires $O(l)$ bits. The chance that a wrong pair is detected in the $l$th round is exactly $1/2^l$. So the expected cost of detection is $\sum_1^{\log k} l/2^l = O(1)$. The number of wrong input-pairs is at most the number of input-pairs, hence $\leq k$. Therefore the expected running time is at most $O(k)$ indeed.\\

The chance that a fixed wrong input-pair is not detected is
exactly $1/2^{\log k} = 1/k$. But if instead of testing in each
round only once, we can test twice. This doubles the bits sent but
halves the chance of erring. The number of bits sent during a
track-back reamins the same. The chance that a fixed wrong
input-pair is not detected becomes $1 - \sum_1^{\log k} 3/4^l =
1/4^{\log k} = 1/k^2$. So the chance that there is any
undiscovered wrong input-pair is at most $k/k^2 = 1/k \leq 1/2$.
Therefore we have proved $R^{0,pub}(\times _k EQ) = O(k)$.
\end{proof}

Note that obviously the complexity is at least $k$ because the answer is $k$ bits. So we know that $R^{pub}(\times _k EQ) = \Theta(k)$.\\

Using a certain random protocol for $GT$ (see \cite{KN96} pp.
170-171) a similar argument shows that $R^{pub}(\times _k GT) =
O(k \cdot R^{pub}(GT))$. This suggests the following conjecture:

\begin{conj} $R^{pub}(\times _k f) = O(k \cdot R^{pub}(f))$
\end{conj}

\clearpage
\section{Complexity Classes}
In the first subsection we follow the book (\cite{KN96} pp. 58-59), the other subsections are completely our results.

\subsection{Classes}
We can categorize communication complexity problems just like computational complexity problems into classes. In fact, we can define all important classes in CC as well. However, because here every problem is solvable in $n + 1$ time, the classes are rather compared to \textit{polylog(n)} than to polinoms of n. The basic classes are:
\begin{itemize}
\item[] $P^{cc} = \{f: D(f) = \polylog (n)\}$, \item[] $NP^{cc} =
\{f: N^1(f) = \polylog (n)\}$, \item[] $coNP^{cc} = \{f: N^0(f) =
\polylog (n)\}$, \item[] $BPP^{cc} = \{f: R(f) = \polylog (n)\}$,
\item[] $RP^{cc} = \{f: R^1(f) = \polylog (n)\}$, \item[]
$coRP^{cc} = \{f: R^0(f) = \polylog (n)\}$.
\end{itemize}

Note that $RP^{pub, cc}=RP^{cc}$ because of Lemma \ref{pub-priv}.\\
Of course these classes are related to sequence of functions for all $n$-s, not for a single function. Unlike in computational complexity, here we can prove or disprove almost all relations among these classes. We summarize the known results below:

\begin{thm} Hierarchy of Communication Complexity Classes:
\item[(0)] $P^{cc} \subseteq RP^{cc} \subseteq NP^{cc}$.
\item[(1)] $P^{cc} = NP^{cc} \cap coNP^{cc}$.
\item[(2)] $P^{cc} \neq NP^{cc} \neq coNP^{cc}$.
\item[(3)] $P^{cc} \neq RP^{cc} \neq coRP^{cc}$.
\item[(4)] $coNP^{cc} \backslash BPP^{cc} \neq \emptyset$.
\end{thm}
\begin{proof}
(0) follows from $D(f) \geq R^1(f) \geq N^1(f)$.\\

(1) is known as the Aho-Ullman-Yannakakis Theorem \cite{AUY83}.\\

For (2) and (3), functions NE and EQ are the counterexamples because $D(EQ) = D(NE) = N^1(EQ) = N^0(NE) = n + 1$, while $N^1(NE) = N^0(EQ) = \log n$, $R^1(NE) = R^0(EQ) = O(\log n)$, $R^1(EQ) = R^0(NE) = \Omega(n)$.\\

For (4), the $DISJ$ function is the counterexample. $N^0(DISJ) \leq \log n$ is trivial while $R(DISJ) = \Omega(n^{1/2})$ was first proved by Babai et al. \cite{BFS86}, then it was improved to $\Omega(n)$ by Kalyanasundaram and Schnitger \cite{KSch87}, later this proof was simplified by Razborov \cite{R90}.
\end{proof}

To study further relations among complexity classes, it is useful to define \textit{reducibility} and \textit{completeness}, as in the book (\cite{KN96} pp. 58-59):

\begin{defi} \label{red} Let the inputsize of $f$ be $n$ and the inputsize of $g$ be $m$ where $m = 2^{\polylog (n)}$. $f$ is reducible to $g$ if a pair of functions $h_x: \{0,1\}^n \rightarrow \{0,1\}^{m}$ and $h_y: \{0,1\}^n \rightarrow \{0,1\}^{m}$ exist such that $f(x,y) = 1 \Leftrightarrow g(h_x(x), h_y(y))=1$. For a class $C$, a series of functions $g$ (the same function with different inputsizes) is $C$-complete, if $g\in C$ and if every $f\in C$ is reducible to $g$ (with the proper inputsize).
\end{defi}

It can be easily proved that $DISJ$ is $coNP^{cc}$-complete. We omit the proof here, because in Section \ref{oracles} we will give an equivalent definition of reducibility and prove this statement there.\\

We can define the analog classes of the polynomial hierarchy:

\begin{itemize}
\item[] $\Sigma_0^{cc}=\{f:$ The communication matrix of $f$ is 1
except a single \textit{rectangle} that is filled with 1s $\}$.
\item[] $\Pi_0^{cc}=co\Sigma_0^{cc}=\{f:$ The communication matrix
of $f$ is 0 except a single \textit{rectangle} that is filled with
0s $\}$. \item[] $\Sigma_{k+1}^{cc} = \{f:
f=\bigvee_{i=1}^{2^{\polylog (n)}}f_i, \: f_i\in \Pi_k^{cc}\}$
\item[] $\Pi_{k+1}^{cc} = co\Sigma_{k+1}^{cc} = \{f:
f=\bigwedge_{i=1}^{2^{\polylog (n)}}f_i, \: f_i\in
\Sigma_k^{cc}\}$
\end{itemize}

It can be easily proved that $\Sigma_1^{cc}=NP^{cc}$ and similarly
$\Pi_1^{cc}=coNP^{cc}$. We know complete problems for all of these
classes (to be discussed in Section \ref{oracles}), although we
still do not know whether $\Sigma_2^{cc}\stackrel{?}{=}
\Pi_2^{cc}$.

\subsection{Space-bounded Communication}
We can even define a corresponding class to PSPACE. To do this, first we have to define the cost of communication measured in Space instead of Time. This class might be useful for proving the existence of special Oracles or maybe even used to give a lower bound for the Space used by a Turing machine computing a certain function, like we can use $EQ$ to give a quadratic lower bound for the Time used to decide the language of \textit{Palindromes} on a one-tape Turing machine (see \cite{D97}).\\

We redefine the communication in the following way: \A and \B are
still supernatural beings capable of computing anything but now
they only have a limited amount of memory and that is common. The
minimum size of this common memory that they can use to evaluate
the given function $f$ shall be denoted by $S(f)$. At the
beginning it is filled with 0s. Then in each step one of the
players can put there an arbitrary message depending only on the
previous message and his input. They are finished when both of
them knows the value of $f(x,y)$. We can also imagine this as two
people communicating who have no memory at all (however, they can
remember their own input) and are allowed to send each other a
rewritable disk. The question is how big the disk has to be if
both of them wants to know the value of $f(x,y)$. Let's see an
example.

\begin{claim} \label{SEQ} $S(EQ) = \log(n) + O(1)$.
\end{claim}
\begin{proof}
We present a construction. \A sends her bits one after the other along with their ordinal number and a leading 1, meaning that it is up to \B to speak. \B replies to each message with his bit with the same ordinal number and a leading 0. This requires $2 + \log n$ space. If in a step his bit differs from her, they know that the answer is 0, the algorithm is over. If they finish sending all their bits, the answer is 1. Therefore, $S(EQ) \leq \log n + O(1)$.
\end{proof}

To get lower bounds for $S(f)$, we need the following lemma:

\begin{lem} \label{DexpS} $S(f) \geq \lambda(D(f)) \geq \log D(f) - \log \log D(f)$.
\end{lem}
\begin{proof}
It is sufficient to show that if we have a protocol using $S(f)$ space, then we can construct an another one using $S(f)2^{S(f)}$ time. The new protocol is the following: \A sends in order all her replies to all possible messages. After this \B can easily simulate the old protocol by himself. So this takes $S(f)2^{S(f)}$ time, just what we wanted.
\end{proof}

Combining methods like the one presented in Claim \ref{SEQ} with Lemma \ref{DexpS} we get:

\begin{cor}
\item[] $S(EQ) = \Theta(\log n)$.
\item[] $S(GT) = \Theta(\log n)$.
\item[] $S(IP) = \Theta(\log n)$.
\item[] $S(DISJ) = \Theta(\log n)$.
\end{cor}

Moreover, in fact we have proved $S(f)/\log n \stackrel{n \rightarrow \infty}{\longrightarrow} 1$ for all of the above four functions.\\

Note that in the proof of Lemma \ref{DexpS} we used only that \A
has no memory. In the case when \B can remember everything, we
denote the required space instead of $S(f)$ by $S^A(f)$. Obviously
$S(f) \geq S^A(f)$. The corollary of Lemma \ref{DexpS}:

\begin{cor} $S^A(f) \geq \lambda(D(f))$.
\end{cor}

But we can improve the upper bound for $S^A$:

\begin{claim} $S^A(f)2^{S^A(f)} \leq n + O(1)$.
\end{claim}
\begin{proof}
We have to give a protocol that is using only $b$ space where $b2^b = n$. We divide up the $n$ bits of \A into $n/b$ disjoint blocks, each of size $b$. In each step \B sends the name of a block and \A sends back that block of hers. At the end \B knows everything, he can compute the function, this requires $\max (b, \log n/b) = b$ space, just what we wanted.
\end{proof}

Combining the previous corollary with the previous claim we get:

\begin{cor}
\item[] $S^A(EQ)2^{S^A(EQ)} = n + O(1)$.
\item[] $S^A(GT)2^{S^A(GT)} = n + O(1)$.
\item[] $S^A(IP)2^{S^A(IP)} = n + O(1)$.
\item[] $S^A(DISJ)2^{S^A(DISJ)} = n + O(1)$.
\end{cor}

The exact value of $S(EQ)$ is yet unknown.\\

These results suggest to define the following classes:
\begin{itemize}
\item[] $SPACE^{cc}(g(n)) = \{f: S(f) = g(n)\}$ \item[]
$PSPACE^{cc} = \bigcup_{k=1}^\infty SPACE^{cc}(\log^k(n)) = \{f:
S(f) = \polylog (n)\}$
\end{itemize}

So far we were unable to find complete problems, however, we
believe that there exists some.

\subsection{Oracles} \label{oracles}
We can even introduce \textit{Oracles} in communication
complexity. Doing so, we have to mix the fact that they both know
only their own inputs and yet asking the Oracle, it should respond
some useful information. To achieve this, at each question they
ask simultaneously; whether they raise a question or just
communicate, is built into the protocol and their questions depend
only on the previous communication and their own input. The Oracle
is a function $g(x',y')$ whose inputsize is $m$ where $m = 2^{\polylog (n)}$. When they ask a question, \A determines
$x'$ from her input and the previous communication and similarly
\B determines $y'$. The Oracle's answer is $g(x',y')$. Each
question counts to be 1 bit of communication. (Or a space on the
tape in the Space-bounded case.)
\\

Eg., let the Oracle accept inputs of length $n$ and be able to say $GT(x',y')$ (whether its first input is bigger then the second one with the usual ordering). Now \A and \B can compute $EQ$ in 2 steps. First they input $x$ and $y$, then $\overline{x}$ and $\overline{y}$. If both answers are 1, then $x=y$, otherwise not. We denote this by $D^{GT}(EQ) = 2$.\\

Another example: The Oracle accepts inputs of length $2n$ and is able to say $DISJ(x',y')$. Now they can solve the problem with a single question. \A inputs $x\overline{x}$ and \B inputs $\overline{y}y$. This implies $D^{DISJ}(EQ) = 1$.\\


Now we give an equivalent definition to reducibility (Def \ref{red}):

\begin{claim} Let the inputsize of $f$ be $n$ and the inputsize of $g$ be $m$ where $m = 2^{\polylog (n)}$. $f$ is reducible to $g$ iff $D^g(f) = 1$.\\
(Unless $g$ is constant, further on we are going to assume
this.)
\end{claim}
\begin{proof}
If $f$ is reducible to $g$, then they ask $h_x(x)$ and $h_y(y)$. The Oracle's answer will equal to $f(x,y)$.\\
If $D^g(f) = 1$, it means that they communicated only a single bit or asked only one question. The former means that $D(f) = 1$, hence it is reducible to any $g$ that is not constant. The latter means that they ask a question that depends only on their inputs, thus they have an $h_x$ and an $h_y$ function, just what we wanted to prove.
\end{proof}

\begin{cor} \label{red1} Note that if $D^g(f) = 1$, then $f(x,y) = g(h_x(x),h_y(y))$.
\end{cor}

\begin{claim} \label{DISJ compl} $\forall f\in \Pi_1^{cc}$ \: $D^{DISJ}(f) = 1$.
\end{claim}
\begin{proof}
The 0 entries of $f$ can be covered by $m = 2^{\polylog (n)}$
0-rectangles. \A and \B both input a $m$ bit long message to the
Oracle indicating which 0-rectangles intersect their inputs. If
these are not disjoint, then the value of the function is
obviously 0. If these are disjoint, then the value cannot be 0,
because all 0s are contained in one of the $m$ rectangles, thus
the answer is 1.
\end{proof}

\begin{cor} $DISJ$ is $coNP^{cc}$-complete.
\end{cor}

Now we are going to give $\Sigma_k^{cc}$-complete and
$\Pi_k^{cc}$-complete functions for all $k$. In fact we are going
to show that if we have a given class $C$ and a $C$-complete
function, how we can construct a complete function for $\exists C
= \{f: f = \bigvee_{i=1}^{2^{\polylog (n)}}f_i, \: f_i\in C\}$.
Let $coDISJf: \{0,1\}^{mn} \times \{0,1\}^{mn} \rightarrow
\{0,1\}$ be the following function: $coDISJf((x_1 \ldots x_m),(y_1 \ldots y_m)) = 1 \Leftrightarrow \exists i \: f(x_i, y_i) =
1$.

\begin{claim} If $g$ is $C$-complete, then $coDISJg$ is $\exists C$-complete.
\end{claim}
\begin{proof}
The proof is very similar to Claim \ref{DISJ compl}. Let $f \in
\exists C$ be an arbitrary function, the inputs are $x$ and $y$.
By the definition of $\exists C$, the set of the 1 entries of $g$
is the union of the 1 entries of $m = 2^{\polylog (n)}$
functions: $f_1, \ldots , f_m$, $\forall i f_i \in C$, thus
$D^g(f_i) = 1$. This means because of Corollary \ref{red1} that
for suitable $x_i, y_i$ input-pairs $f(x,y) = g(x_i,y_i)$. Now
they ask from the Oracle: $(x_1 \ldots x_m)$ and $(y_1 \ldots
y_m)$. If the answer is 1, then there is an $i$ such that
$g(x_i,y_i) = 1$, therefore $f(x,y) = 1$ as well. If the answer is
0, then $\forall i \: g(x_i,y_i) = 0$, this implies $f(x,y) = 0$,
so this solves our problem indeed.
\end{proof}

This gives complete functions for $\Sigma_k^{cc}$ and $\Pi_k^{cc}$
for every $k$. Eg., $coDISJDISJ$ is $\Sigma_2^{cc}$-complete. It
is the easiest to imagine the inputs as two $n \times n$ matrices
and the question is whether the two inputs have a disjoint row.
However, it still remains an open question whether the classes
$\Sigma_2^{cc}$ and $\Pi_2^{cc}$ are the same or not.

\end{document}